\documentclass[aps,prb,longbibliography,floatfix,reprint]{revtex4-1}
\usepackage{amsmath}
\usepackage{longtable}
\usepackage{amsfonts}
\usepackage{graphicx} 
\usepackage{comment}
\usepackage{color}
\usepackage{soul,xcolor}
\setstcolor{red}   

\begin{document}

\title{Heat pulse propagation and nonlocal phonon heat transport in
    one-dimensional harmonic chains}

\author{Philip B. Allen}
\email{philip.allen@stonybrook.edu}
\affiliation{Department of Physics and Astronomy, Stony Brook University, Stony Brook, NY 11794-3800, USA}

\author{Nhat A. Nghiem}
\affiliation{Department of Physics and Astronomy, Stony Brook University, Stony Brook, NY 11794-3800, USA}

\date{\today}


\begin{abstract}
Phonons are the main heat carriers in semiconductor devices.  In small devices, heat is not driven
by a local temperature gradient, but by local points of heat input and removal.  This
complicates theoretical modeling.   Study of the
propagation of vibrational energy from an initial localized pulse
provides insight into nonlocal phonon heat transport.
We report simulations of pulse propagation in one dimension.
The 1d case has tricky anomalies, but provides the simplest pictures
of the evolution from initially ballistic toward longer time diffusive propagation.  Our results show
surprising details, such as diverse results from different definitions of atomistic
local energy, and failure to exhibit pure diffusion at long times.
Boltzmann phonon gas theory, including external energy insertion, is applied to this inherently 
time-dependent and nonlocal problem.   
The solution, using relaxation time approximation for impurity scattering,
does not closely agree with the simulated results.

\end{abstract}

\maketitle


\section{Introduction}

When heat is inserted in a local region of a semiconductor, phonons (labeled by $Q=(\vec{q},j)$,
wavevector and branch) carry heat ballistically until they scatter.  The mean free path $\ell_Q$
of different phonons is very diverse.  At large distances from the source, the local heat
current $j(r,t)$ is no longer ballistic.  The propagation of current beomes increasingly diffusive.  
Modelling of the crossover from ballistic to diffusive is difficult.
It can be illuminated by simulations of simple model cases.
The one-dimensional chain is the simplest model, with only one branch of phonons.  In spite of
the worry of oversimplifying the problem, the clarity of one-dimensional pictures provides some
useful insights.

\par
\begin{figure}
\includegraphics[angle=0,width=0.4\textwidth]{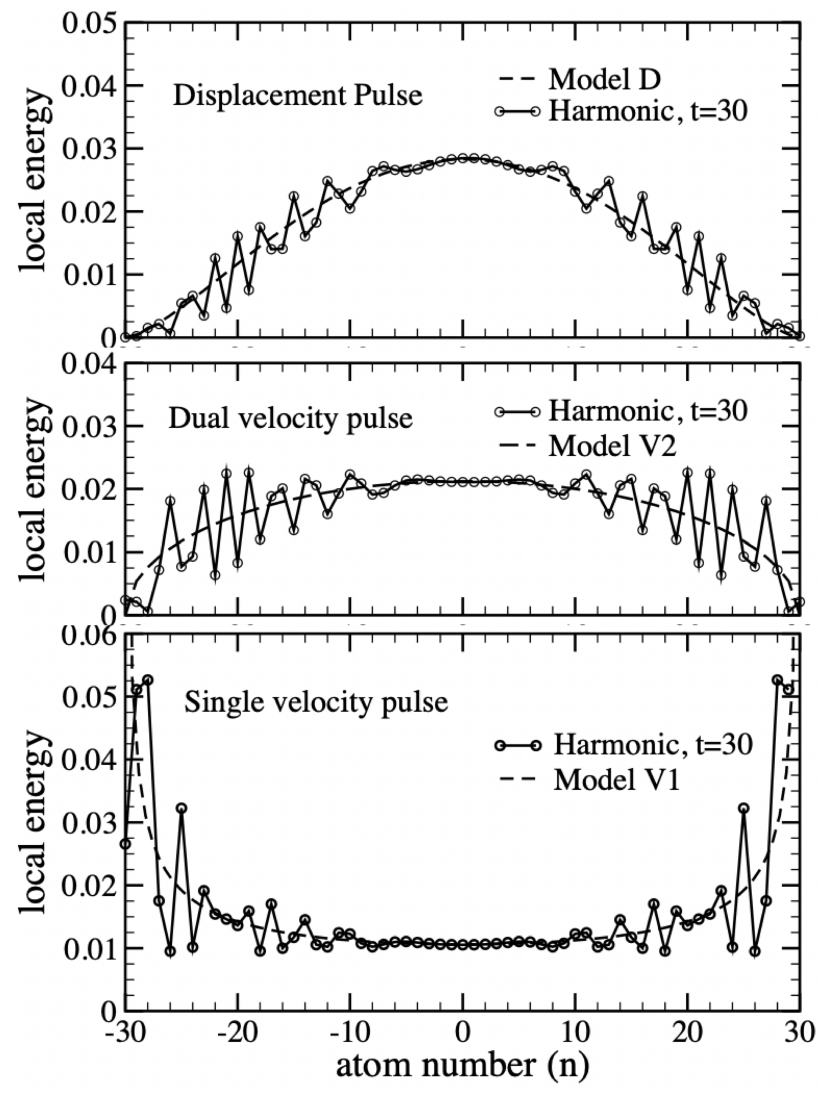}
\caption{\label{fig:3pulses} Circles connected by solid lines are the computed local energy
$E_a(\ell)$ at $t=30$ for three types of pulses (see Eq. \ref{eq:Eell} and Table \ref{table:I}).  All three pulses
evolve from energy inserted at $t=0$ and $r\sim 0$ into a harmonic chain.  
The D and V2 pulses are initiated on atoms 0 and 1, but their
positions on the horizontal axis are shifted by -1/2 to make them symmetric around $r=0$.
The dashed curves are models derived from a continuum picture, such as Boltzmann theory, and 
given in Eqs. \ref{eq:p1}, \ref{eq:p2}, \ref{eq:p3}.}
\end{figure}
\par

Figure \ref{fig:3pulses} shows the local energy at atom sites on a one-dimensional chain of $N=200$ atoms 
coupled by nearest-neighbor harmonic springs.  The system has periodic boundary conditions.  At time $t=0^-$, the 
chain is in its ground state.  At $t=0$, one or two central atoms are
disturbed, starting a pulse.  No further external disturbance is applied.  The classical time
evolution is computed from Newton's laws.  This is the simple example analyzed in this paper.
In section II, the chain and the pulses are defined.  In section III, the definition of ``local energy" 
is analyzed and found to be more interesting than expected.  In section IV, a continuum description
is chosen and shown to explain the different results shown in Fig. \ref{fig:3pulses}.  In section V, the
Boltzmann equation is written for the nonlocal time-dependent problem, and its collisionless limit
is shown to agree with the previously chosen continuum description.  In section VI, simulations are repeated
for an ensemble of mass-disordered chains, showing evolution in time in a direction toward diffusive
energy propagation.  In section VII, ``pure'' diffusion is defined, and shown to have imperfect ability
to accurately explain heat propagation at long times.  Section VIII returns to the Boltzmann description,
with scattering from mass disorder added.  The one-dimensional chain presents difficulties in
the perturbative description of such scattering.  The relaxation-time approximation (RTA) gives
a sensible-looking approximate formula but is found not to agree accurately with the simulated
energy propagation.  Section IX gives a brief presentation of the effect of anharmonic scattering
on the 1-d pulses.  Section X summarizes the conclusions.

The main points of this paper are: 
(1) To clarify the ideas of local heat and local temperature. 
(2)To study the ``crossover'' from the ballistic propagation of Fig. \ref{fig:3pulses} toward diffusive
heat propagation when phonons start to evolve toward a local equilibrium $T(r,t)$ because of 
scattering \cite{Liu2014,VermeerschI2015}.  
(3) To test the form and the accuracy of a non-local Boltzmann equation description \cite{Hua2020}.

\section{The harmonic linear chain}

The chain has atoms of mass $M=1$ separated by distance
$a=1$ and connected by springs of constant $K=1$.
The harmonic normal modes have frequency $\omega_Q=\omega_M \sin(Qa/2)$ where
$\omega_M=2\sqrt{K/M}=2$.  Time is measured in units $\sqrt{M/K}=2/\omega_M=1$.
The modes propagate at velocities $v_Q=v_M \cos(Qa/2) {\rm sign}(Q)$
where $v_M=\sqrt(K/M)a=1$.  The unit of energy is $E_0=1=Ka^2=Mv_M^2=M\omega_M^2 a^2/4$.
The leading edges of the pulse propagate at the
speed of sound, $\pm v_M=\pm 1$, as is seen at $t=30$ in Fig. \ref{fig:3pulses}.

The Hamiltonian of the chain is
\begin{equation}
\mathcal{H}=\sum_\ell \left[ \frac{P_\ell^2}{2M}+\frac{1}{2}K (u_\ell - u_{\ell+1})^2\right].
\label{eq:}
\end{equation}
Atoms have displacements $u_\ell$ around average positions $r_\ell=\ell a$.
The general solution of Newton's equations of motion is
\begin{equation}
u_\ell(t)=\sqrt{\frac{1}{N}}  \sum_Q A_Q \cos(Q\ell a-\omega_Q t +\phi_Q),
\label{eq:u}
\end{equation}
There are $2N$ free parameters, amplitude $A_Q\ge 0$ and phase $\phi_Q$ for each
normal mode $Q$.  The wavevector has the form $Q=(2\pi/a) (n/N)$, and the integer $n$
lies in the range $-N/2 < n \le N/2$.  The pulses shown in 
Fig. \ref{fig:3pulses}, labeled D, V2, and V1, are generated by initial disturbances given in Table \ref{table:I}.  
The ``V1'' (or ``velocity'') pulse has only the
central atom given a velocity $v_0$ at $t=0$.  The ``V2'' (or ``dual velocity'') pulse
has two central atoms given equal and opposite velocities.  
The ``D'' (or ``displacement'') pulse has two
central atoms given equal and opposite displacements at $t=0$. 

The first result to notice is the interesting diversity of pulse shapes
for different initial disturbances, as was first noticed in ref. \onlinecite{Hua2017}.  
The first issue to resolve is whether local energy is well-defined.

\begin{table}
\begin{center}
\begin{tabular}{ ccccc } 
 \hline
  \ name \  & \ \ \ $\Delta u_0$ \ \ \ & \ \ \ $\Delta u_1$ \ \ \ & \ \ \ $\Delta v_0$ \ \ \ &  \ \ \ $\Delta v_1$ \\ 
 \hline
 V1 & 0 & 0 & $\sqrt 2$ & 0 \\
 V2 & 0 & 0 & -1 & 1 \\  
 D & $-1/\sqrt 3$ & $1/\sqrt 3$ & 0 & 0 \\
 \hline
\end{tabular}
\caption{\label{table:I}Properties of pulses.  The shift $\Delta u$ (in units of $a$) of initial displacement, or 
$\Delta v$ (in units of $v_M$) of initial velocity, is scaled so that the new coordinates ($u_0+\Delta u_0, \ u_1+\Delta u_1, 
 \ v_0+\Delta v_0, \ v_1+\Delta v_1$), of atoms $\ell=0$ and $1$, have total extra energy $E_{\rm pulse}=1$.
 The values in the table give $E_{\rm pulse}=0$ at $T=0$, {\it i.e.} when $u_\ell(t)=0$ for $t<0$}
\end{center}
\end{table}

\section{Defining local energy}

\begin{figure}
\includegraphics[angle=0,width=0.4\textwidth]{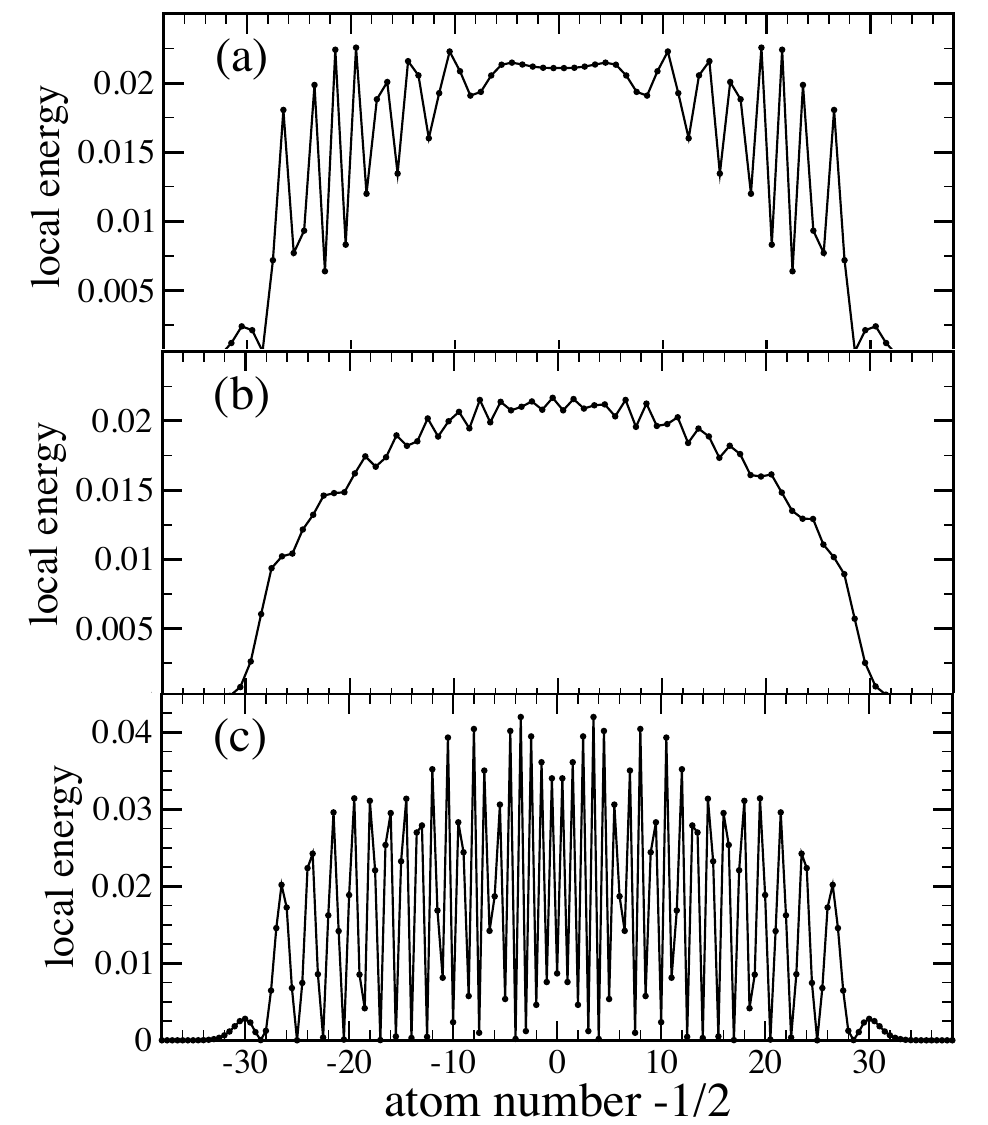}
\caption{\label{fig:3types} The harmonic V2 pulse at temperature $T=0$ and time $t=30$, 
using three ways (Eqs. \ref{eq:Eell}-\ref{eq:Eellc})
of assigning potential energy to a local site energy, $E(\ell)$.  In (a) and (b), the pulse energy is 
$E_{\rm pulse}=E_0=\sum_\ell E(\ell)$.
The area under the curves is $E_0$, chosen to be 1.  In (c), however, the energy has two parts, 
$\sum_\ell [KE(\ell)+PE(\ell + 1/2)]$.
The values in the graph have been arbitrarily doubled so that the total energy, also 1, is the area under the graph.
These graphs demonstrate that the ``continuum picture" of energy density (the dashed curves in Fig. \ref{fig:3pulses}) is 
more sensible than these local pictures.}
\end{figure}

Local energy $E(\ell,t)$ cannot be unambiguously defined \cite{Hardy1963}.  In classical physics
it should obey $\sum_\ell E(\ell,t)=\mathcal{H}$.  The version plotted in 
Fig. \ref{fig:3pulses} is defined as
\begin{eqnarray}
E_a(\ell,t)&=&\frac{Mv_\ell(t)^2}{2} \nonumber \\
&+& \frac{K}{4}\left[(u_{\ell-1}(t)-u_\ell(t))^2  + (u_\ell(t) - u_{\ell+1}(t))^2\right]. \nonumber \\
\label{eq:Eell}
\end{eqnarray}
Each atom is assigned its own kinetic energy, and half of the potential energy of the
springs to its left and right.  This is a commonly used definition, but
is not unique.  Two other sensible choices for distributing potential energy to different sites are
\begin{equation}
E_b(\ell)=\frac{Mv_\ell^2}{2} + \frac{K}{2}\left[(2u_\ell^2  - u_\ell(u_{\ell-1} + u_{\ell+1})\right],
\label{eq:Eellb}
\end{equation}
\begin{equation}
E_c(\ell)=\frac{Mv_\ell^2}{2} \ {\rm and} \  E_c(\ell+1/2)= \frac{V}{2} (u_\ell - u_{\ell+1})^2.
\label{eq:Eellc}
\end{equation}
The conventional version $E_a$ will be used in this paper, but it is interesting to see how it
compares with versions $E_b$ and $E_c$, as shown in Fig. \ref{fig:3types}.  
The large oscillations seen in version (c) are surprising, and the smoothness seen in version 
(b) is even more surprising.  
This diversity illustrates our second result,
a well-known fact: local energy at the atom level is not a clear concept.  However, if local energy is averaged 
over a few nearby atoms, it becomes less diverse.  
The ambiguity also exists in quantum treatments.  Marcolongo {\it et al.} \cite{Marcolongo2016}
and Ercole {\it et al.} \cite{Ercole2016} have shown how this ambiguity does not affect the computation
of bulk transport.  In the next section we will see 
that local energy makes more sense in a continuum theory than an atomistic theory.

\par
\par
\par
\begin{figure*}[!ht] \centering
\includegraphics[angle=0,width=0.85\textwidth]{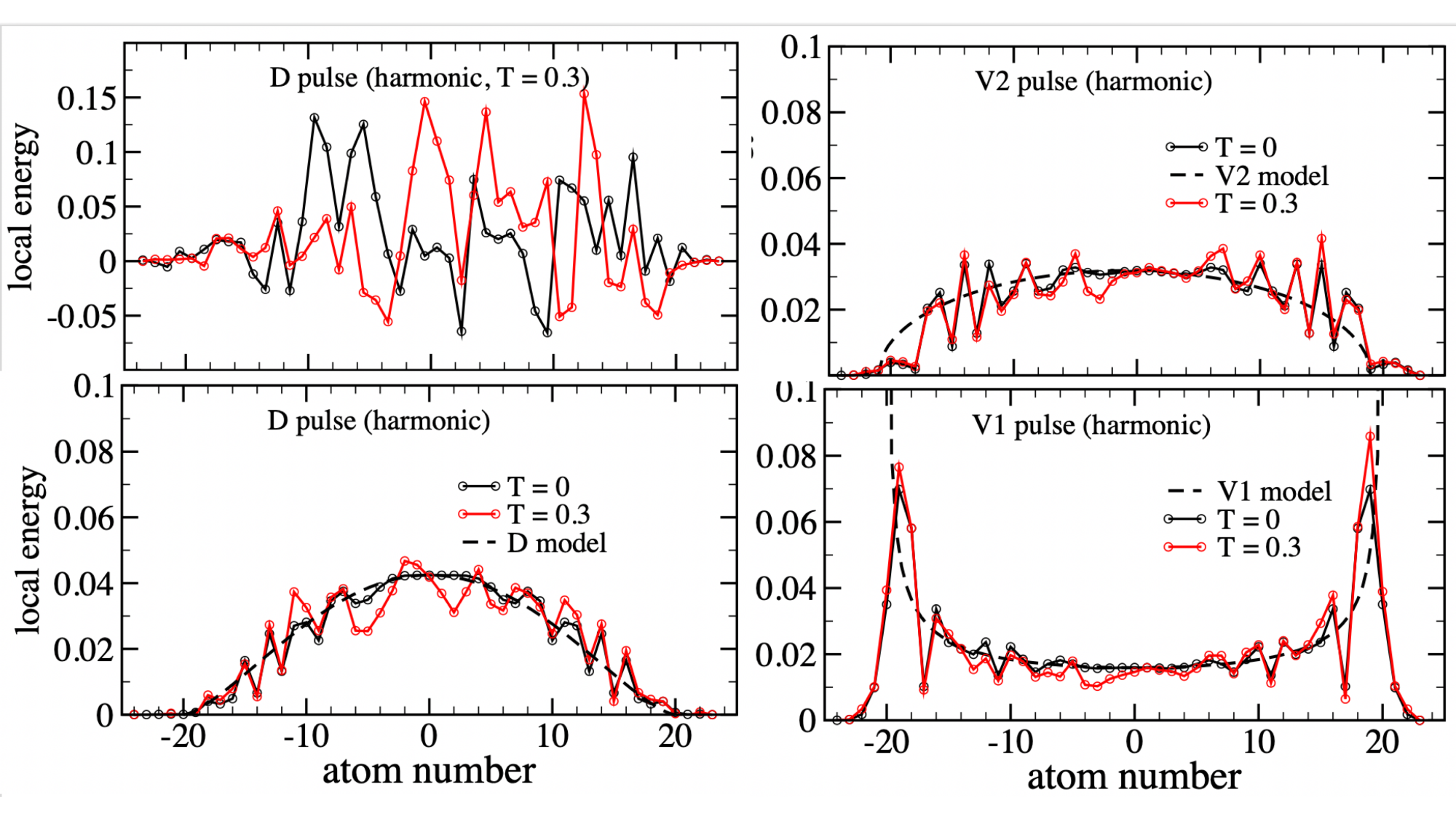}
\caption{\label{fig:kappa}  Pulses of total energy $\sum_\ell \Delta E_a(\ell) = E_0 = 1$ at $t=20$, 
inserted at $t=0$ to harmonic chains
whose random positions and velocites at $t=0^-$ are
thermalized at temperature $T=0.3$ ($E_{\rm tot,th}=0.3N$).  The local energies $\Delta E_a(\ell)$ shown
in the graphs are the differences between $E_a(\ell,t=20)$ with and without pulse insertion. 
The first graph shows D pulses inserted with two typical initial conditions.  The
other graphs show, in red, the results averaged over 1000-member ensembles with random thermal initial conditions.
The black results repeat the zero-temperature pulses shown in Fig. \ref{fig:3pulses}.  The finite $T$ pulse shapes,
after ensemble averaging, are converging towards the zero-temperature pulse shapes.}
\label{fig:Tnot0}
\end{figure*}
\par

Figure \ref{fig:3pulses} shows  $E_a(\ell,t)$ for pulses, inserted at $t=0$
into zero-temperature ({\it i.e.} stationary, $T=0$) chains.
Figure \ref{fig:Tnot0} shows the same pulse forms, at $t=20$, inserted into chains with a pre-existing
thermal distribution of velocities and displacements.  In $T>0$ cases, the initial pulse amplitudes 
($\Delta u_{0,1}$ or $\Delta v_{0,1}$) are scaled
from those in table I to make the total extra energy $\sum_\ell \Delta E_a(\ell)$ of all pulses of the ensemble equal to 1.
The pulse profiles in Figs. \ref{fig:3pulses} and \ref{fig:Tnot0} were computed in two different ways:  (1) by numerical integration
of Newton's laws, and (2), by finding the coefficients $A_Q$ and $\phi_Q$  in Eq. \ref{eq:u}.  These coefficients
are independent of time, and can be found if the positions $u_\ell$
and velocities $v_\ell = du_\ell/dt$ are known at any chosen time:  
\begin{equation}
A_Q e^{i(\phi_Q -\omega_Q t)} =\sqrt{\frac{1}{N}} \sum_\ell \left[u_\ell(t)+i\frac{v_\ell(t)}{\omega_Q}\right] e^{-iQ\ell a}.
\label{eq:Aphi}
\end{equation}
For the $T=0$ case, values of $A_Q$ and $\phi_Q$ are given in table \ref{table:II}.  They are derived 
from the $t=0^+$ positions and velocities shown in table \ref{table:I} at $t=0$.  

\begin{table}
\begin{center}
\begin{tabular}{ c|c|c|c } 
 \hline
  & Amplitude $A_Q$ & phase & modal energy \\ 
name   & from Eq. \ref{eq:Aphi} & $\phi_Q$ & $\Delta E_Q^{\rm ext}$ \\  
 \hline
 V1 &  $ \frac{\Delta v_0}{\sqrt{N}\omega_Q}$ & $\frac{\pi}{2}$ & $\frac{E_0}{N}$ \\ 
 V2 &  $\frac{2\Delta v_0}{\sqrt{N}\omega_Q} \sin\left(\frac{Qa}{2}\right)$ 
 & $\pi - \frac{Qa}{2}$ & $\frac{2E_0}{N} \sin^2\left(\frac{Qa}{2}\right)$ \\ 
  D &   $ \frac{2\Delta u_0}{\sqrt{N}}\sin\left(\frac{Qa}{2}\right)$  
  & $-\frac{\pi}{2} -\frac{Qa}{2}$ & $\frac{8E_0}{3N} \sin^4\left(\frac{Qa}{2}\right)$ \\  
 \hline
\end{tabular}
\caption{\label{table:II} More properties of pulses: the distribution among normal modes $Q$ of the phonon
amplitude $A_Q$, phase $\phi_Q$, and energy, for the pulses of table \ref{table:I} inserted at $T=0$.}
\end{center}
\end{table}

The chains are harmonic, so the pulses propagate ballistically.  The left or right parts have root mean square (rms) 
displacements $\bar{r}$ defined as
\begin{equation}
\bar{r}(t) \equiv \left[\frac{\sum_\ell (\ell a)^2 \Delta E(\ell,t)}{\sum_\ell \Delta E(\ell,t)}\right]^{1/2}.
\label{eq:xbar}
\end{equation}
The rms displacements increase at speeds $v_{\rm rms}=d\bar{r}/dt \approx  v_M/\sqrt{n}$,
with $n=2, \ 4, \ 6$ for the V1, V2, and D pulses, respectively.  
These values are derived from the continuum description described next.

\section{A continuum description}

The energy content of each normal mode is
\begin{equation}
E(Q)=\frac{1}{2}M\omega_Q^2 A_Q^2.
\label{eq:EQ}
\end{equation}
Using values of  $A_Q$ from Table \ref{table:II}, the mode energies are also shown in Table \ref{table:II}.
A continuum description uses spatially averaged atomic coordinates, and requires a new definition of 
local energy density $E(r,t)$.  An appropriate definition for a pulse originating at $(r,t)=(0,0)$ is
\begin{equation}
\Delta E(r,t)=\sum_Q \Delta E_Q^{\rm ext} \delta(r-v_Q t).
\label{eq:Ext}
\end{equation}
After integrating over $Q$, the results for the three pulses are  
\begin{equation}
\Delta E_{V1}(r,t)= \frac{E_0 }{\pi v_M t} \left(1-\left(\frac{r}{v_Mt}\right)^2 \right)^{-1/2} \theta(v_Mt-|r|),
\label{eq:p1}
\end{equation}
\begin{equation}
\Delta E_{V2}(r,t)= \frac{2E_0 }{\pi v_M t} \left(1-\left(\frac{r}{v_Mt}\right)^2 \right)^{1/2} \theta(v_Mt-|r|),
\label{eq:p2}
\end{equation}
\begin{equation}
\Delta E_D(r,t)= \frac{8E_0 }{3\pi v_M t} \left(1-\left(\frac{r}{v_Mt}\right)^2 \right)^{3/2} \theta(v_Mt-|r|).
\label{eq:p3}
\end{equation}
The total energy $\int dr \Delta E(r,t)$ is $E_0$ in all three cases.  These formulas are shown as dashed lines
in Figs. \ref{fig:3pulses} and \ref{fig:Tnot0}.  The continuum model agrees well with an average of nearby values of the 
local atomic energy $E(\ell)$ for all of the three pulse types $E_\alpha(\ell)$.
The rms centers of energy of the propagating pulses are then 
\begin{equation}
\bar{r}_{\rm continuum}(t) \equiv \left[\frac{\int dr r^2 \Delta E(r,t)}{\int dr \Delta E(r,t)}\right]^{1/2}.
\label{eq:xbarcont}
\end{equation}
This is where the result $d\bar{r}/dt \rightarrow v_M/\sqrt{n}$ (with $n = 2, \ 4, \  6$) came from.

It had been our original guess that when the pulse propagates in a thermal background, the fine structure
in $E(\ell)$ would disappear and the result would resemble the continuum version $E(r,t)$.  
The computation in Fig. \ref{fig:Tnot0} shows that this guess was wrong.  The fine structure remains.
A proof that this should happen is given in the Supplemental Material \cite{suppl}.

\section{Non-local Boltzmann equation, collisionless}

The formulas given in Eqs. \ref{eq:p1}, \ref{eq:p2}, \ref{eq:p3} came from a sensible hypothesis, Eq. \ref{eq:Ext},
which will now be derived from Boltzmann theory.  Pulse behavior is
fundamentally non-local, {\it i.e.} not determined by the local deviation of temperature $T(r,t)$ from background $T_0$. 
The usual local Boltzmann equation is very successful \cite{Lindsay2019} in describing the bulk thermal conductivity $\kappa$, appearing in 
the Fourier law $j=-\kappa\nabla T$, where the temperature gradient $\nabla T$ is constant in space and time, or varying
slowly on the scale of phonon mean free paths $\ell_Q$ and lifetimes $\tau_Q$.  Pulse propagation
requires an extension of the usual local version.
 For at least 60 years \cite{Simons1960,Guyer1966,Mahan1988} there have been developments of Boltzmann theory 
 aimed at studying systems driven
 at small distance scales, {\it i.e.} distances comparable to quasiparticle mean free paths.  There are many recent
 studies, for example, refs. 
 \onlinecite{Johnson2013,Koh2014,Hua2014,Hua2014b,Hu2015,Hua2017,Hua2018,Hua2019,Hua2020,Hua2020b}.
 There are also models outside of Boltzmann theory that work even 
 when geometries are too complex \cite{Beardo2021} for the Boltzmann method.

The Peierls Boltzmann equation \cite{Peierls1929} (PBE) uses quantum wave/particle duality to describe the system as a
gas of phonon particles in a continuous space with coordinate $r$, rather than
discrete sites $r_\ell=\ell a$ on a lattice.  Phonons can be treated either as classical or quantum particles.
A correct treatment of Boltzmann theory with the full scattering 
operator $(\partial N_Q /\partial t)_{\rm coll}$ gives the same low frequency 
(and spatially homogeneous) transport properties as a
self-consistent treatment by Green-Kubo theory to second order in interactions \cite{Horie1964,Gotze1967,Klein1969,Ranninger1967,Spohn2006}.  
The fundamental object is
$N_Q(r,t)$, the occupancy per unit volume $V$ (where $V$ = length $L$ in 1d) of phonon mode $Q$ at $(r,t)$. 
The spatial sum $\int dr N_Q(r,t)=N_Q(t)$ is the mode occupancy.  The PBE is
\begin{equation}
\frac{\partial N_Q}{\partial t}=- v_Q\frac{\partial N_Q}{\partial r}+
 \left(\frac{\partial N_Q}{\partial t}\right)_{\rm coll}
+\left(\frac{\partial N_Q}{\partial t}\right)_{\rm ext}.
\label{eq:PBE}
\end{equation}
 Local energy is 
$E(r,t)=(1/V)\sum_Q \hbar\omega_Q N_Q(r,t)$.  Heat current density is 
$j(r,t)=(1/V)\sum_Q \hbar\omega_Q v_Q N_Q(r,t)$.  Local energy is conserved.  Summing Eq. \ref{eq:PBE} 
over $Q$ (after multiplying by $\hbar\omega_Q$) gives
\begin{equation}
\frac{\partial E(r,t)}{\partial t}= -\nabla j(r,t) + \left(\frac{\partial E(r,t)}{\partial t} \right)_{\rm ext}.
\label{eq:Econs}
\end{equation}
This holds because collisions conserve energy locally, $\sum_Q \hbar\omega_Q (\partial N_Q /\partial t)_{\rm coll}=0$.
We will use the quantum version, with an equilibrium Bose-Einstein 
distribution $N_Q\rightarrow n_Q(T_0)=[\exp(\hbar\omega_Q/k_B T_0)-1]^{-1}$, and take
the classical limit $n_Q(T) = k_B T /\hbar\omega_Q$ when comparing with simulations.

The scattering (or collision) term $(\partial N_Q/\partial t)_{\rm coll}$ tries to drive the local distribution
to a local thermal distribution.  Writing the distribution as $n_Q(T(r,t))+\Phi_Q(r,t)$, where $\Phi_Q$ is 
the deviation from local equilbrium, and linearizing, Eq. \ref{eq:PBE} becomes
\begin{eqnarray}
\frac{\partial n_Q}{\partial T}\frac{\partial T}{\partial t}+\frac{\partial \Phi_Q}{\partial t}&=&
-v_Q\left( \frac{\partial n_Q}{\partial T} \frac{\partial T}{\partial r} +\frac{\partial \Phi_Q}{\partial r} \right) \nonumber \\
&& -\sum_{Q^\prime}C_{QQ^\prime}\Phi_{Q^\prime}+\left(\frac{\partial N_Q}{\partial t}\right)_{\rm ext}.
\label{eq:PBElin}
\end{eqnarray}
where $-C_{QQ^\prime}$ is the linearized scattering operator.
The atoms are driven by external manipulation that changes the 
Newtonian state $\{u_\ell,v_\ell\}$ to $\{u_\ell+\Delta u_\ell,v_\ell+\Delta v_\ell\}$ at $t=0$, for $\ell=0$ and $1$.  
It changes the phonon amplitudes and phases to give the starting pulse shape.  
A continuum version of the change must be created by the term 
$(\partial N_Q/\partial t)_{\rm ext}$ in the Boltzmann equation.
External driving $(\partial N_Q/\partial t)_{\rm ext}$ has only recently appeared in phonon
Boltzmann theory \cite{Hua2014b,Vermeersch2014,VermeerschI2015,Hua2015}; its 
form and significance is still open to discussion.
Boltzmann theory does not deal directly with amplitudes $A_Q$.  These are indirectly included {\it via} the mode energy
$MA_Q^2 \omega_Q^2 /2 \rightarrow \hbar\omega_Q (N_Q+1/2)$.  Coherent phase relations $\phi_Q$ between different
quasiparticles $Q$ cannot be handled.
A choice for the external term $(\partial N_Q/\partial t)_{\rm ext}$ driving the
distribution function $N_Q$ away from equilibrium is
\begin{equation}
\left( \frac{\partial N_Q}{\partial t} \right)_{\rm ext}=\frac{\Delta E_Q^{\rm ext}}{\hbar\omega_Q} \delta(r)\delta(t).
\label{eq:Ndotext}
\end{equation}
A very similar version of Boltzmann theory applied to time-domain thermoreflectance was given in ref. \onlinecite{VermeerschI2015}.
The energy inserted by the pulse into mode $Q$ is $\hbar\omega_Q \times \Delta N_Q^{\rm ext}$,
where $\Delta N_Q^{\rm ext}=\int dr \int dt (\partial N_Q(r,t)/\partial t)_{\rm ext}$.
The total pulse energy given to the system is clearly correct:
\begin{equation}
E_{\rm pulse}^{\rm Boltzmann}= \sum_Q \Delta E_Q^{\rm ext}.
\label{eq:Epulse0}
\end{equation}

Because of linearity and periodic boundary conditions, it is convenient
to Fourier transform to $N_Q(k,\omega)$,
\begin{equation}
N_Q(k,\omega)=\frac{1}{L}\int_{-L/2}^{L/2} dr \int_{-\infty}^\infty dt N_Q(r,t) e^{-i(kr-\omega t)}.
\label{eq:}
\end{equation}
Equation \ref{eq:PBElin} becomes
\begin{eqnarray}
\sum_{Q^\prime}C_{QQ^\prime}\Phi_{Q^\prime} &+& i(kv_Q-\omega)\Phi_Q =\nonumber \\
&-& i(kv_Q-\omega)\frac{\partial n_Q}{\partial T}\Delta T(k,\omega) + \frac{\Delta E_Q^{\rm ext}}{\hbar\omega_Q}. \nonumber \\
\label{eq:PBEkw}
\end{eqnarray}
In the harmonic case, there are no collisions and $C_{QQ^\prime}=0$.  The solution in Fourier space is
\begin{equation}
\frac{\partial n_Q}{\partial T}\Delta T(k,\omega) + \Phi_Q(k,\omega) =
\frac{\Delta E_Q^{\rm ext}/\hbar\omega_Q}{i(kv_Q-\omega-i\eta)}.
\label{eq:harsol}
\end{equation}
The left hand side is $\Delta N_Q(k,\omega)=N_Q(k,\omega)-n_Q(T_0)$.
Transforming back to $(r,t)$-space,
\begin{eqnarray}
\Delta N_Q(r,t) &=& \frac{L}{2\pi}\int dk \int\frac{d\omega}{2\pi} 
\frac{\Delta E_Q^{\rm ext}/\hbar\omega_Q}{-i(\omega +i\eta-kv_Q)}e^{i(kr-\omega t)} \nonumber \\
&=& \frac{\Delta E_Q^{\rm ext}}{\hbar\omega_Q} \delta(r-v_Q t)
\label{eq:Bsoln}
\end{eqnarray}
%
The local energy density is 
\begin{equation}
\bar{E}_{\rm pulse}(r,t) = \sum_Q \hbar\omega_Q \Delta N_Q(r,t) = \sum_Q \Delta E_Q^{\rm ext}\delta(r-v_Q t).
\label{eq:ExtB}
\end{equation}
This agrees exactly with Eq. \ref{eq:Ext}.  These results provide confidence in the
insertion term added to the Peierls Boltzmann equation.  We also learn that, in the continuum
description, harmonic pulse shapes (Eqs. \ref{eq:p1}-\ref{eq:p3}) are independent of $T$, because the
temperature $T_0$ in the Boltzmann treatment did not have to be specified.  The less obvious result,
that harmonic pulse shapes in the atomistic version are also independent of $T$, 
is explained in the Supplemental Material \cite{suppl}.

\section{Mass disorder}

\par
\begin{figure*}
\includegraphics[angle=0,width=0.95\textwidth]{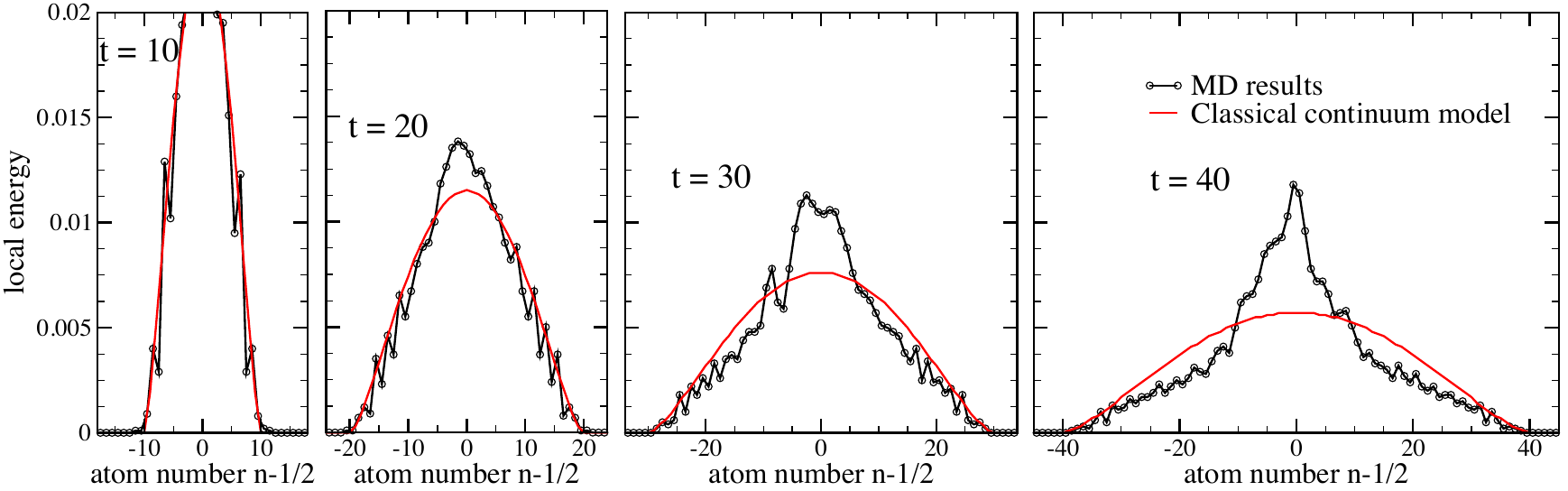}
\caption{\label{fig:disorder} Total energy profile $E_a(\ell)$ of a {\bf D}-pulse at $T=0$, after spreading 
in the lattice with 10\% of the masses increased by 1.5.  The profiles are averaged over
100 random realizations of mass disorder.  Time is in units $\sqrt{M/K}$. 
The red curves are the ballistic prediction of Eq. \ref{eq:p3}.    
The value of $E_0$ has been rescaled from 1 to 0.27, corresponding to $u_0=0.3$, which
is the value chosen for this and subsequent computations.}
\end{figure*}
\par

Disorder adds an interesting complication to heat conduction in
1-d harmonic crystals \cite{Matsuda1970,Visscher1971}, namely Anderson localization \cite{Anderson1958}. 
In disordered metals of dimension 2 or less, ignoring electron-electron interactions, all single-particle electron eigenstates
are localized \cite{Abrahams1979}.  At $T=0$, an electron inserted into a localized state
cannot propagate.  Localization of phonons is similar \cite{Sheng2005,Kirkpatrick1985,Luckyanova2018}, except that
small $Q$ acoustic phonons have to travel very long distances before localization appears \cite{Lepri2003}.  
Reference \onlinecite{Allen1998} gives
the example of a wave packet progating on a weakly mass-disordered chain.  Ballistic propagation is
seen at short distances and times, diffusive propagation at intermediate ones, and Anderson localization at long 
distances and times.  When $T>0$, interactions with phonons allow a localized electron to hop to neighboring
localized states, which causes slow diffusion.  
Anharmonic interactions have a similar effect \cite{Fabian1996} on localized phonons in insulators.
If disorder is not too great, phonon quasiparticles are a realistic model at intermediate times and distances.
Ballistic phonons eventually scatter from disorder and evolve toward diffusive at moderate to long distances
and times, before localizing.  A perturbative treatment of scattering can likely describe the 
evolution before localization sets in.

We now add mass defects to allow deviation from ballistic propagation, by 
randomly choosing 10\% of the atoms, and increasing their masses from $M=1$ to $M^\ast=1.5$.  
Results at various times for a D pulse are shown in Fig. \ref{fig:disorder}.  The lattice is
still harmonic, but the Hamiltonian is no longer diagonal in the plane-wave basis, Eq. \ref{eq:u}.  Before
ensemble averaging, the pulse shape varies depending on the locations of the mass defects
relative to the point of pulse insertion.  The {\bf D}-pulse shapes 
of Fig. \ref{fig:disorder} have been averaged over 100 different random placements of the 
altered masses.

The pulse shape at $t=10 \sqrt{M/K}$ is not much altered from pure ballistic behavior.  The pulse has
propagated only a distance of $\pm 10$ atoms, and encountered typically only two impurities.
As time proceeds, there is increasing deviation from the ballistic pulse shape predicted in Eq. \ref{eq:p3}, and
shown in the red curves.  By $t = 40$, the 
fraction of the energy at distances $< 10a$ has failed to diminish the way ballistic propagation does.  The energy in the
intermediate $10 - 30$ atoms has diminished more than ballistic propagation does.  The crossover toward diffusion 
is underway.   The disorder is weak, so the Anderson localization lengths $\xi_Q$ are mostly
longer than the propagation distance ($\le 40a$) studied here.

\section{Pure Diffusion}
\label{sec:purediff}

The opposite extreme from ballistic propagation is pure diffusion.   The equation
for the energy $\Delta E$ propagating diffusively from a pulse $P$ is
\begin{equation}
\left(\frac{\partial}{\partial t}-D\frac{d^2}{dr^2} \right)\Delta E(r,t)=P(r,t)=E_0\delta(r)\delta(t)
\label{eq:diff}
\end{equation}
This follows from energy conservation, Eq. \ref{eq:Econs}, provided there is a local relation $j=-\kappa dT/dr$
between current $j(r,t)$ and temperature $T(r,t)$.  It also uses as the definition of temperature $\Delta E(r,t)= C\Delta T(r,t)$.
Then the diffusion constant is $D=\kappa/C$ where $\kappa$ and $C$ are bulk thermal conductivity and specific heat.
The solution of Eq. \ref{eq:diff} is
\begin{equation}
\Delta E_{\rm diff}(r,t)=\frac{E_0}{\sqrt{4\pi Dt}} e^{-r^2/4Dt}\theta(t),
\label{eq:purediff}
\end{equation}
where $\theta(t)$ is the Heaviside function.

Pure diffusion is inconsistent with a quasiparticle picture of pulse evolution.  One argument is that it violates the
rule that lattice vibrational energy cannot propagate faster than the speed of sound $v_M$.  It is not
necessarily a large violation.  An estimate of the size uses $\kappa\approx Cv\ell=Cv^2 \tau$.  
The diffusive exponent $r^2/4Dt$ is then
approximately $(t/4\tau)(r/v_M t)^2$ (where $\ell$ and $\tau$ are rough measures of
mean free path and lifetime of phonons).  Therefore, when
$t=4\tau$, there is some diffusive energy at $r>v_M t$, which decays rapidly as $r/v_M t$ or $t/4\tau$ increases.

\begin{figure}
\includegraphics[angle=0,width=0.45\textwidth]{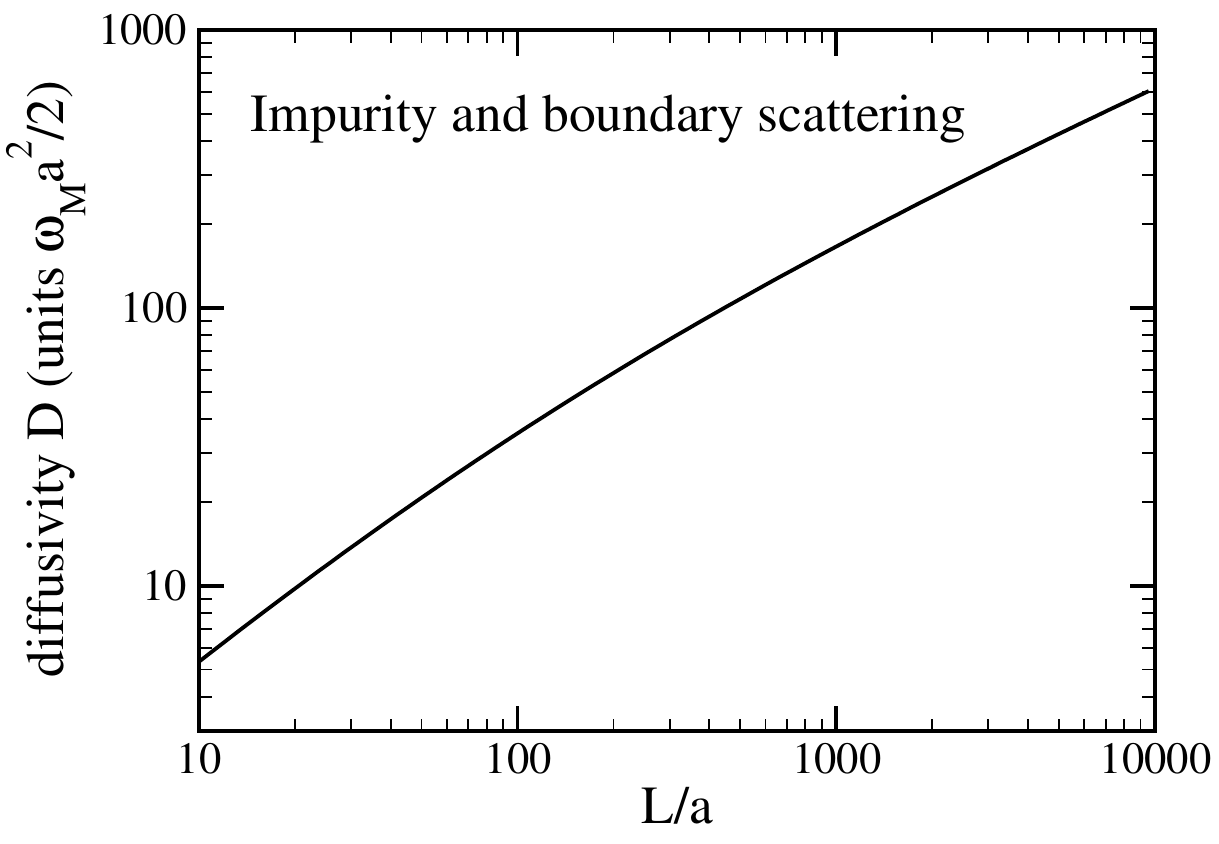}
\caption{\label{fig:DvsL} The diffusivity in bulk RTA theory, 
from Eq. \ref{eq:DwL} and \ref{eq:DwL2}, using Eq. \ref{eq:tauQ} for the disorder scattering,
with the dimensionless strength $\epsilon=1/90$.  The diffusivity diverges as $\sqrt(L/\epsilon)$ as $L\rightarrow\infty$.}
\end{figure}

Another argument for the inapplicability of pure diffusion to quasiparticles is that mean free paths of small $Q$
acoustic phonons typically diverge as a power, $\ell_Q \propto 1/Q^p$, causing $D$ to diverge.
This is a correct result, not an error of perturbation theory.  
A formula for $D=\kappa/C$ is found from the standard RTA solution of Boltzmann theory for $\kappa$ in the bulk limit,
\begin{equation}
D=\frac{\sum_Q C_Q v_Q^2 \tau_Q}{\sum_Q C_Q} \rightarrow \frac{1}{N} \sum_Q v_Q^2 \tau_Q.
\label{eq:Ddiv}
\end{equation}
The second form is the classical limit where $C_Q=\hbar\omega_Q dn_Q/dT \rightarrow k_B$.
When $1/\tau_Q$ arises from mass disorder, the small
$Q$ scattering rate $1/\tau_Q$ in $d=1$ goes as $(Q^2)$  \cite{Matsuda1970,Li2014,Rubin1971,Verheggen1979,Dhar2008}, 
shown explicitly in appendix A.  The divergence is not limited to one dimension.  
Small $Q$ scattering from mass defects is closely analogous to Rayleigh scattering of light from
density fluctuations.  In 3d, both light and phonon scattering have $Q^4$ dependences at small $Q$.
The $Q$-sum needed to compute $D$ or $\kappa$ diverges as $1/Q^2$ in both $d=1$ and $d=3$, 
unless another scattering process adds a term to $1/\tau_Q$ that blocks the divergence. 
When $T$ is small enough that mean free paths $\ell_Q=v_Q \tau_Q$ reach sample dimensions $L$, 
the zero denominator from
the diverging defect term becomes non-zero due to boundary scattering $L=v_Q \tau_{\rm boundary}$. 
Glassbrenner and Slack analyze experiments which illustrate this \cite{Glassbrenner1964}.  It
is seen in clean but isotopically disordered crystals \cite{Thacher1967,Asen-Palmer1997,Lindsay2013}.
Eqn. \ref{eq:Ddiv} can be replaced by 
\begin{equation}
D=\sum_Q v_Q^2 \left[(1/\tau_Q)_{\rm disorder}+|v_Q|/L \right]^{-1}.
\label{eq:DwL}
\end{equation}
The sum now converges, diverging for large $L$ as $D\propto \sqrt{L}$ in $d=1$.  
The $\sqrt{L}$ scaling of $\kappa$ in d=1 was noticed earlier \cite{Matsuda1970,Keller1978}. 
The resulting $D$ is plotted {\it versus} $L/a$ in Fig. \ref{fig:DvsL}.  Our Newtonian simulation on a loop
of 200 atoms has no boundary scattering because of periodic boundary conditions.  If we had used instead
rough boundaries, then $L$ in Eq. \ref{eq:DwL} would be $100a$, resulting in $D=35\omega_M a^2/2$
(see Appendix A).

\par
\par

\begin{figure}
\includegraphics[angle=0,width=0.45\textwidth]{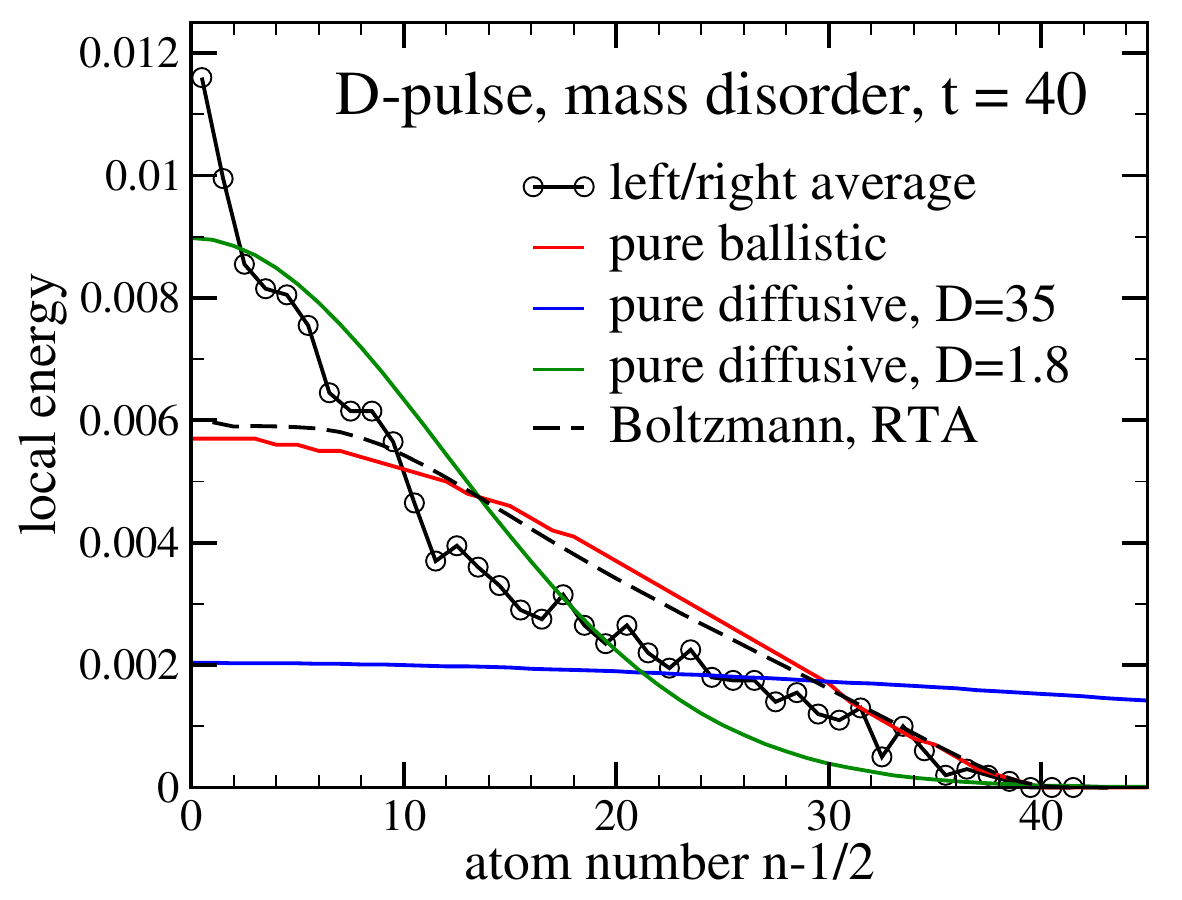}
\caption{\label{fig:t40} The same $t=40$ pulse energy profile $E_a(\ell)$ as in Fig. \ref{fig:disorder},
but showing only $x>0$ and averaging the left and right propagating portions. 
The red curve is the ballistic prediction of Eq. \ref{eq:p3}, and the blue curve is the diffusive prediction of Eq. \ref{eq:purediff},
with diffusion constant $D=35 a^2 \omega_M /2$ as explained in the text.  The green curve is the same,
except $D=1.8 a^2 \omega_M /2$ is used to make diffusion appear closer to the computed pulse profile.  
The dashed curve is
the Boltzmann prediction (with no boundary scattering) using version {\bf (1)} of RTA, 
shown in Eq. \ref{eq:2versions} 
and explained in section \ref{sec:BMD}.  All ``theoretical'' curves are scaled to $E_0=0.27$ 
to agree with the numerical simulation. }
\end{figure}

In spite of the inapplicability of Eq. \ref{eq:purediff}, it is interesting to compare it to the numerical results.
Figure \ref{fig:t40} compares the simulation result of the disordered {\bf D}-pulse at $t=40$
(the last graph of Fig. \ref{fig:disorder}) with the formulas for pure ballistic
and pure diffusive propagation.  For the majority of atoms (all but atoms 10-14)
the computed pulse energy is closer to the green curve illustrating pure diffusion (with $D=1.8$) than to the
red curve of pure ballistic behavior.  However, the agreement with pure diffusion is poor,
and more important, the choice $D=1.8$ was chosen to give a curve for $\Delta E(r,t=40)$ reasonably
close to the simulation, but it is totally unrealistic, corresponding to a nanoscale sample with boundaries
at $L \approx \pm 3a$, as seen from Fig. \ref{fig:DvsL}.

\par
\par

\section{Boltzmann theory with mass disorder}
\label{sec:BMD}

How does non-local Boltzmann theory treat pulse shape evolution as altered by disorder?  Here
we answer this question within the relaxation time approximation (RTA), and find that the
results have the correct trend but do not agree closely with simulations.  The collision operator $C_{QQ^\prime}$
in Eq. \ref{eq:PBEkw} is replaced in RTA by $1/\tau_Q \delta_{QQ^\prime}$, where $1/\tau_Q$ is
the diagonal element, $C_{QQ}$, 
\begin{equation}
1/\tau_Q=\epsilon\omega_M\frac{\sin^2(Qa/2)}{\cos(Qa/2)}.
\label{eq:tauQ}
\end{equation}
This is derived in appendix A, Eq. \ref{eq:tauQ2}.
The RTA solution is a modified version of Eq. \ref{eq:harsol},
\begin{equation}
\Phi_Q(k,\omega) =
\frac{-i(kv_Q-\omega)\frac{\partial n_Q}{\partial T}\Delta T(k,\omega) + \Delta E_Q^{\rm ext}/\hbar\omega_Q}
{1/\tau_Q+i(kv_Q-\omega)}.
\label{eq:RTAsol}
\end{equation}
This gives $N$ equations (one for each mode $Q$), for the $N$ unknown functions $\Phi_Q(k,\omega)$.  There is an additional unknown,
the local temperature shift $\Delta T(k,\omega)$.  An extra equation,
to supplement Eq. \ref{eq:RTAsol}, is needed.

The requirement of energy conservation says that
$\sum_Q\hbar\omega_Q \left( d\Phi_Q/dt\right)_{\rm coll}=0$.
The correct scattering operator satisfies this automatically, but the RTA version,
$(dE/dt)_{\rm coll,RTA}=-\sum_Q \hbar\omega_Q \Phi_Q/\tau_Q=0$, is {\bf not} automatically satisfied.  
Forcing $\Phi_Q$ to satisfy this equation as well as the linearized PBE is one way to obtain
the extra equation needed to determine $\Delta T(k,\omega)$ and $\Delta T(r,t)$.  This definition of temperature
which we call ``version {\bf (1)}'', is
not a normal one.  A possible alternative is to define temperature in terms of the local energy,
 $E(r,t)=\sum_Q \hbar\omega_Q n_Q(T(r,t))$ 
or $E(k,\omega)=\sum_Q \hbar\omega_Q (dn_Q/dT)\Delta T(k,\omega)$.   
This, called ``version {\bf (2)}'',
is the definition of $\Delta T(k,\omega)$ used in the full Boltzmann equation with the correct
scattering operator.  It requires the deviation $\Phi_Q(r,t)$ to have no net energy.  
This definition of temperature is quite normal - it is sensible in a quasiparticle
theory, although perhaps not demanded by nonequilibrium thermodynamics.

The two possible versions of the extra equation, needed in RTA, are then
\begin{equation}
\bf{(1)} \ \sum_Q \frac{\hbar\omega_Q \Phi_{Q,{\rm RTA}}}{\tau_Q}=0; \ {\rm or} \ {\bf (2)} \ 
\sum_Q \hbar\omega_Q \Phi_{Q,{\rm RTA}} =0. 
\label{eq:2versions}
\end{equation}
We find that version {\bf (1)} gives a more realistic answer, in agreement with previous
numerical \cite{Allen2018} and theoretical 
\cite{Vermeersch2014,VermeerschI2015,Hua2015} work.

\par
\begin{figure}
\includegraphics[angle=0,width=0.45\textwidth]{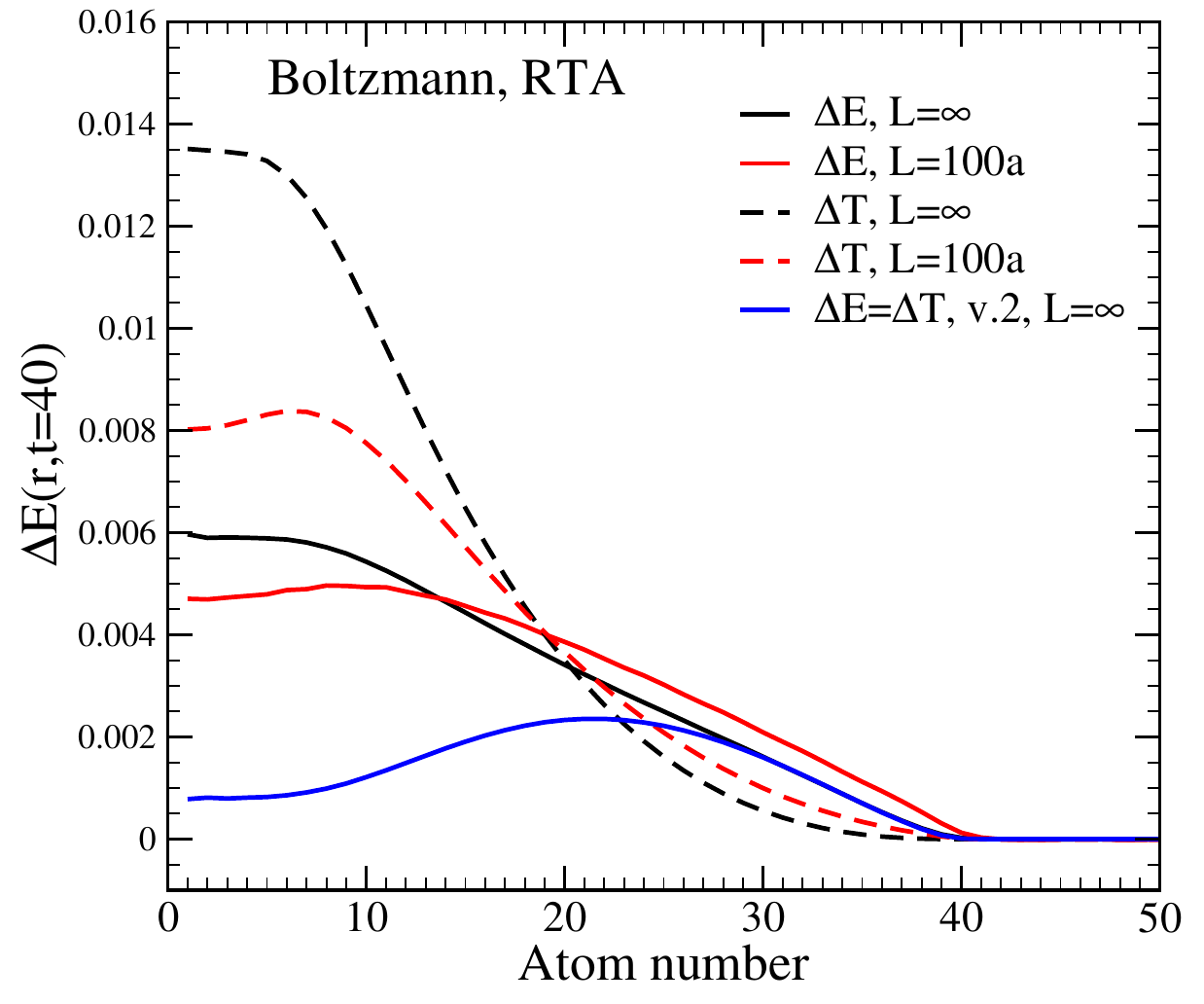}
\caption{\label{fig:RTA} RTA predictions for the total energy profile $E_a(\ell)$ of a {\bf D}-pulse at
$t=40$ and $T=0$, after spreading in the lattice with 10\% of the masses increased by 1.5.  
The input energy $E_0$ is set to 0.27, for comparison with numerical results.
The red and black curves use version {\bf (1)} of Eq. \ref{eq:2versions}, and the blue curve uses version {\bf (2)}. 
The total energy (solid curves) correctly agrees with $E_0$ in version {\bf (1)}, but in version {\bf (2)}, is smaller by a factor 0.415.  In version {\bf (1)}, the
thermal energy is (incorrectly, we believe) larger than $E_0$ by a factor 1.58.  }
\end{figure}
\par
\par
\begin{figure}
\includegraphics[angle=0,width=0.45\textwidth]{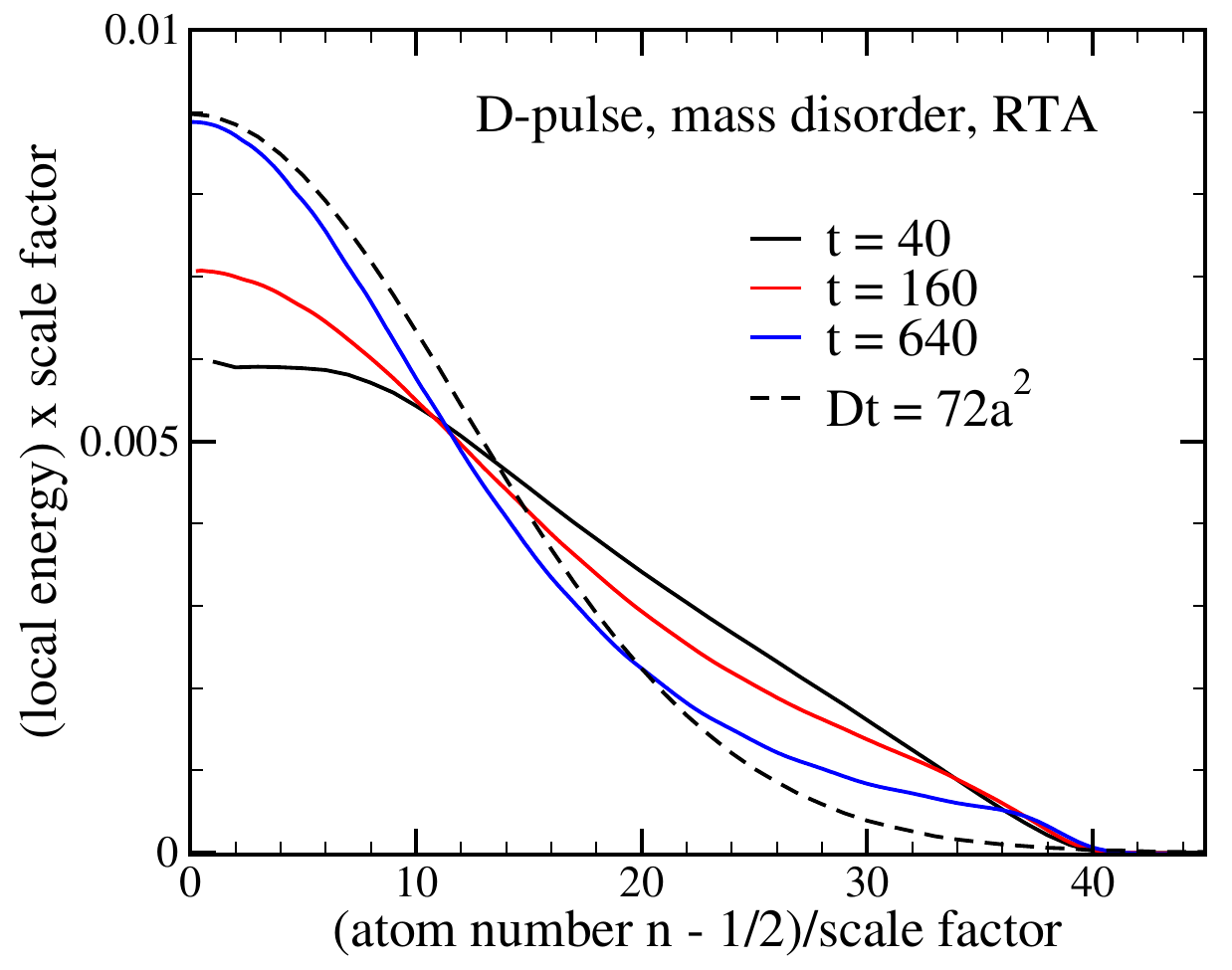}
\caption{\label{fig:RTAlarget} Boltzmann results using RTA, with impurity but no boundary scattering, at three times, 
$t =$ 40, 160, and 640.  Time is
in units $2/\omega_M=1$.  The scale factors 1, 4, and 16 are used to keep the correct area ($E_0=0.27$) while scaling
the horizontal distance $v_M t$ to 40 to make the results fit on a single graph.  The dashed
line is pure diffusion $\propto \exp(-x^2/4Dt)$ (Eq. \ref{eq:purediff}) at $t=40$, with the same arbitrary
diffusivity $D=1.8\omega_M a^2 /2$ used in Fig. \ref{fig:t40}, chosen just to illustrate pure diffusion.   }
\end{figure}
\par

The pulse energy predicted by version {\bf (1)} is shown in Fig. \ref{fig:t40}.
It deviates less from ballistic than does the simulation, but in the correct direction.
Energy profiles predicted by both versions of the RTA Boltzmann theory are in Fig. \ref{fig:RTA} for the {\bf D} pulse at $t=40$. 
Details of the computational proceedure are given in
appendix B and in the Supplemental Material \cite{suppl}.
Version {\bf (1)} of RTA theory correctly keeps the total pulse energy equal to $E_0$ as $t$ increases, 
while version {\bf (2)} does not.
Version {\bf (2)} has another weakness, namely the predicted shape of the evolving pulse in Fig. \ref{fig:RTA} 
is very different from the simulated results in Fig. \ref{fig:t40},
unlike version {\bf (1)} where the Boltzmann-RTA pulse shape is sensible.
However, version {\bf (1)} has a temperature profile $\Delta T(r,t)$ in Fig. \ref{fig:RTA}
 that differs from $\Delta E(r,t)$ without physical justification, unlike
version {\bf (2)} which correctly equates $k_B T(r,t)$ and $\Delta E(r,t)$.

Figure \ref{fig:RTAlarget} shows version {\bf (1)} RTA results for long times to illustrate a slow approach toward 
something related to, but not closely agreeing with, pure diffusion, Eq. \ref{eq:purediff}.   
The RTA diffusivity $D=1.8$ used in the dashed curve of Fig. \ref{fig:RTAlarget} uses an unrealistically small
sample size, $\approx 6a$,as mentioned already.  These results confirm the arguments
in Sec. \ref{sec:purediff} that pure diffusion disagrees with quasiparticle theories.

Why do the Boltzmann-RTA results not agree well with the numerical pulse shapes?
The reason could be a mixture of two problems: 
(1) inadequacy of second order Fermi golden rule, related to incipient Anderson localization; and
(2) use of RTA instead of correct local energy-conserving scattering.

\section{anharmonic}

The famous work of Fermi, Pasta, Ulam (FPU) and Tsingou (FPUT) \cite{FPU1955} found unexpected anomalies
in the opposite limit of our pulse propagation, namely the zero temperature behavior of the lowest harmonic
normal mode of the fixed-end chain, when forces differ from harmonic by 
one or the other of two choices for the anharmonic nearest neighbor coupling, 
\begin{equation}
{\mathcal H}_3^\prime = \frac{V_3}{3}\sum_\ell (r_{\ell+1}-r_\ell)^3 \ \  (V_3=\alpha K, \ {\rm the} \ \alpha-{\rm model});
\label{eq:alpha}
\end{equation}
\begin{equation}
{\mathcal H}_4^\prime = \frac{V_4}{4}\sum_\ell (r_{\ell+1}-r_\ell)^4 \ \  (V_4=\beta K, \ {\rm the} \ \beta-{\rm model}).
\label{eq:beta}
\end{equation}

Heat conduction in weakly anharmonic linear lattices has been studied \cite{Lepri2000} 
and reviewed \cite{Lukkarinen2016}.  Recent work
\cite{Dematteis2020,Lepri2020,Wang2020}    seems to confirm (at least qualitatively)
that perturbation theory works in the regime studied here. If so, the perturbative picture says that
mean free paths will diminish as T increases \cite{Maradudin1962,Conway1965}. A crossover to diffusive behavior, 
which FPU indicates might not happen at T = 0, may well happen at $T > 0$, at rates that increase as T increases.

\begin{figure*}
\includegraphics[angle=0,width=0.9\textwidth]{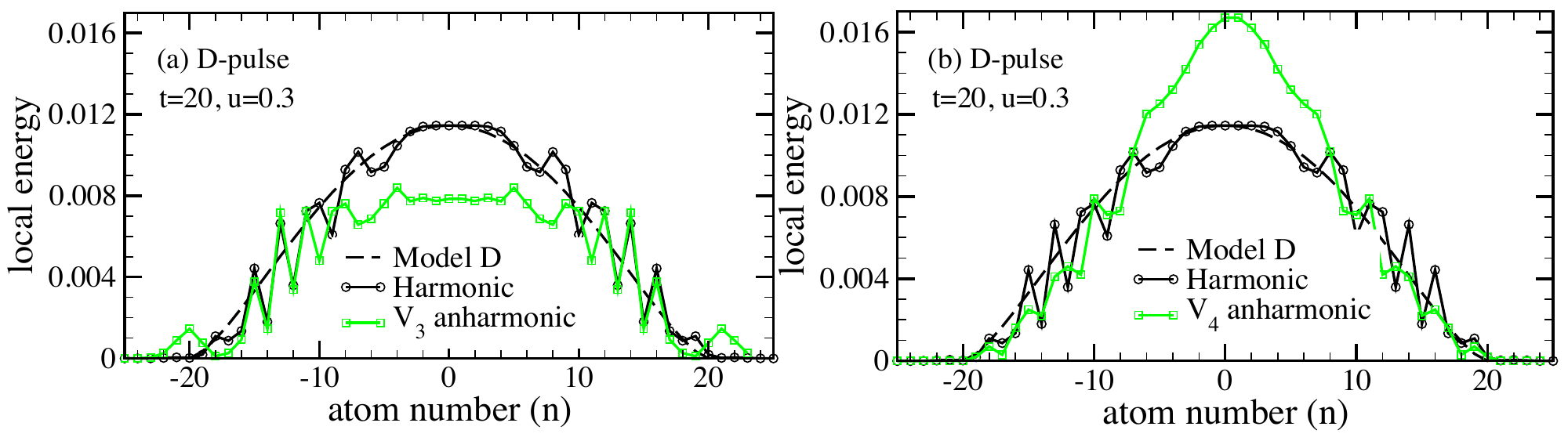}
\caption{\label{fig:anh} Displacement ({\bf D}) pulse at time $t=20$ after
insertion at $T=0$ and $t=0$.  The harmonic ($K_2$)
force constant is 1.  (a) has $V_3=1$ and $V_4=0$;
(b) has $V_3=0$ and $V_4=1$.  
The pure harmonic result ($V_3=V_4=0$) is shown for comparison in both panels.
To see the effect of anharmonicity, initial pulse amplitudes on atoms 0 and 1 were given large
values, 0.3 and -0.3, respectively.  The initial anharmonic energy in (a) is $-0.27 \ \times$ the initial
harmonic energy ($0.27E_0$), and in (b) $+0.12 \ \times$ the initial harmonic energy ($0.27E_0$). }
\end{figure*}

We have looked at $T=0$ anharmonic D-pulse propagation, using both third-order 
($V_3$, also known as ``FPU$-\alpha$'') 
and fourth-order ($V_4$, also known as ``FPU$-\beta$") anharmonicity.  The coefficients
$\alpha$ and $\beta$ in Eqs. \ref{eq:alpha} and \ref{eq:beta} were set to 1.  Local energy $E(\ell,t)$ is defined as
in Eq. \ref{eq:Eell}: the total (harmonic and anharmonic) potential energy of a spring is assigned half to
each neighboring atom.  The results
are in Fig. \ref{fig:anh}.  The initial pulse ($u_{0,1}=\pm 0.3$) is the same as in previous computations, except
the total energy is no longer $0.27E_0$ as in previous computations, because there is additional anharmonic energy,
$-27\%$ in $V_3$ and $+12\%$ in $V_4$.  The initial pulse has the same harmonic amplitudes $A_Q$ and phases
$\phi_Q$ as previously, but anharmonic terms alter these fairly quickly.
As the pulse spreads and $\Delta E(\ell,t)$ spreads out, anharmonic forces diminish.  Amplitudes and phases evolve less,
becoming more stable.  As $t$ increases, local energy propagation from $(x,t)$ reverts more closely to ballistic.
This is especially noticable in the $V_3$ case, 
where atoms $10-16$ seem little affected by anharmonic effects.  The $V_3$ case might be especially difficult to analyze
perturbatively, because in lowest order and 1d, anharmonic decay is essentially kinematically forbidden, requiring higher order
effects to change $A_Q$ and $\phi_Q$.  It would be interesting to study the anharmonic pulse at finite $T$ where anharmonic
effects do not disappear as the pulse propagates.

\par
\par

\section{conclusions}

Time evolution of pulse energy gives useful pictures and insights into non-local phonon transport.
The main conclusions are: (1) Different forms of pulse insertions give interestingly diverse pulse 
energy shapes.  (2) Atomistic local energy $E(\ell,t)$ is not uniquely defined and has surprisingly
different details when different sensible definitions are used.  (3) A continuum picture works well and
enables simple formulas for ballistic propagation.  (4) A non-local version of Boltzmann theory
for the collisionless phonon gas is very successful, and the phonon insertion term $(\partial N_Q/\partial t)_{\rm ext}$
has an unambiguous form.  (5) Mildly disordered harmonic systems have interesting pulse evolution, but are
not well explained by non-local Boltzmann theory in relaxation time approximation (RTA).  
Ambiguity about temperature definition in RTA is a difficulty; previous ideas are confirmed.   (6) Pure diffusion does
not work at the local level, when phonon quasiparticles are good approximations,  even after long pulse evolution times.  (7) The pulse
evolution of 1d anharmonic chains at $T>0$ deserves further study.

\section{acknowledgements}

We are grateful to the Stony Brook Institute for Advanced Computational Science for use of their computer cluster.

\appendix
\section{Phonon relaxation from mass disorder}

The Fermi golden rule for phonon decay by defects is
$1/\tau_Q = (2\pi/\hbar)\sum_{Q\prime} |V_{QQ^\prime}|^2 \delta(\hbar\omega_Q-\hbar\omega_{Q^\prime})$. 
When the interaction $V_{QQ^\prime}$ arises from mass disorder, the decay rate is
\begin{equation}
\left(\frac{1}{\tau_Q}\right)_{\rm imp} = \frac{\pi}{2\hbar} \frac{N_i}{N}
 \left(\frac{\Delta M}{M^\ast}\right)^2 (\hbar\omega_Q)^2 \mathcal{D}(\omega_Q),
\label{eq:tauQ1}
\end{equation}
where $\Delta M=M^\ast-M$ is the mass difference between impurities ($M^\ast$) and pure ($M$) masses,
and $N_i/N$ is the fraction of impurities.
The density of vibrational states of the ordered harmonic chain is
\begin{equation}
\mathcal{D}(\omega) = \frac{1}{\hbar N}\sum_Q \delta(\omega-\omega_Q)
=\frac{2}{\pi\hbar}\frac{1}{(\omega_M^2 -\omega^2)^{1/2}}.
\label{eq:dos}
\end{equation}
Then the decay rate is 
\begin{equation}
\frac{1}{\tau_Q}= \epsilon \frac{\omega_Q^2}{(\omega_M^2 -\omega_Q^2)^{1/2}}, \ \ 
{\rm where} \ \ \epsilon=\frac{N_i}{N} \left(\frac{\Delta M}{M^\ast}\right)^2,
\label{eq:tauQ2}
\end{equation}
and $\epsilon=1/90$ in our simulations.

The formula \ref{eq:DwL} for boundary-limited diffusivity can be written as 
\begin{equation}
D(L)=\frac{\omega_M a^2}{2}\frac{s}{\pi\epsilon} \int_0^{\pi/2} dx \frac{\cos^3x}{\cos^2x+s\sin^2x},
\label{eq:DwL2}
\end{equation}
where $s=2L\epsilon/a$ and $\omega_M a^2/2=1$ is the unit of diffusivity.  At large $s$,
the diffusivity scales as $\sqrt s/\epsilon\propto \sqrt(L/\epsilon)$. 

\section{Solution of Boltzmann equation in RTA}

Using Eq. \ref{eq:RTAsol} and version {\bf (1)} of Eq. \ref{eq:2versions}, the equation for $\Delta T$ is
\begin{equation}
k_B\Delta T_{(1)}(k,\omega)=\frac{A(k,\omega)}{B(k,\omega)}, 
\label{eq:TAB}
\end{equation}
where
\begin{equation}
A(k,\omega)=\frac{1}{N}\sum_Q \frac{\Delta E_Q^{\rm ext}/\tau_Q}{1/\tau_Q -i(\omega-kv_Q)},
\label{eq:1A}
\end{equation}
\begin{equation}
B(k,\omega) =\frac{1}{N}\sum_Q \frac{1}{\tau_Q} \frac{-i(\omega - kv_Q)\left[\frac{\hbar\omega_Q}{k_B}\frac{\partial n_Q}{\partial T}\right]}
{1/\tau_Q -i(\omega-kv_Q)}.
\label{eq:1B}
\end{equation}
Once $\Delta T_{(1)}(k,\omega)$ is computed, the corresponding local energy $\Delta E_{(1)}(k,\omega)$ can be found and Fourier
transformed to get $\Delta E_{(1)}(r,t)$,
\begin{eqnarray}
\Delta E_{(1)}(k,\omega) &\equiv&\sum_Q \hbar\omega_Q\left[ N_Q - n_Q(T_0)\right] \nonumber \\
&=& C(k,\omega)k_B \Delta T_{(1)}(k,\omega) + D(k,\omega),
\label{eq:1En}
\end{eqnarray}
where
\begin{equation}
C(k,\omega)=\frac{1}{N}\sum_Q \frac{1/\tau_Q\left[\frac{\hbar\omega_Q}{k_B}\frac{\partial n_Q}{\partial T}\right]}{1/\tau_Q -i(\omega-kv_Q)},
\label{eq:1C}
\end{equation}
\begin{equation}
D(k,\omega)=\frac{1}{N}\sum_Q \frac{\Delta E_Q^{\rm ext}}{1/\tau_Q -i(\omega-kv_Q)}.
\label{eq:1D}
\end{equation}
The factor in square brackets ($[ \ ]$) in Eqs. \ref{eq:1B} and \ref{eq:1C} (and later in Eq. \ref{eq:1F})
becomes 1 in the classical limit, which is needed for comparison with the numerical pulse spreading.

To implement version {\bf (2)}, use the simpler condition 
\begin{equation}
\Delta E_{(2)}(k,\omega)=k_B \Delta T_{(2)}(k,\omega).
\label{eq:}
\end{equation}
Using Eq. \ref{eq:RTAsol} and version {\bf (2)} of Eq. \ref{eq:2versions}, the equation for $\Delta E$ is
\begin{equation}
\Delta E_{(2)}(k,\omega)=k_B \Delta T_{(2)}(k,\omega)=\frac{D(k,\omega)}{F(k,\omega)}, 
\label{eq:}
\end{equation}
where, except for omission of factors of $1/\tau_Q$ in the numerator, $D(k,\omega)$ is the same as $A(k,\omega)$ and
$F(k,\omega)$ is the same as $B(k,\omega)$.  To be explicit,
\begin{equation}
F(k,\omega) =\frac{1}{N}\sum_Q \frac{-i(\omega - kv_Q)\left[\frac{\hbar\omega_Q}{k_B}\frac{\partial n_Q}{\partial T}\right]}
{1/\tau_Q -i(\omega-kv_Q)}.
\label{eq:1F}
\end{equation}

Details of the numerical calculations, especially the difficult Fourier transforms needed to get pulse shapes
in $(x,t)$-space, are in the Supplemental Materials \cite{suppl}.

\section{Laplace transform method of Vermeersch {\it et al.}}

Vermeersch {\it et al.} (Ref. \onlinecite{VermeerschI2015}) used mixed Fourier (for space variables $r \leftrightarrow k$)
and Laplace (for time variables $t \leftrightarrow \omega$) transforms to solve for the case of a V1 pulse in one dimension.
They used version {\bf (1)} of Eq. \ref{eq:2versions}, but did not transform from $(k,\omega)$ to $(r,t)$.  However, they found
interesting identities in Laplace space, which we will here pursue for arbitary $E_Q^{\rm ext}$, not just the V1 choice $E_Q^{\rm ext}=E_0/N$.
The basis for their identities are the Fourier equations
\begin{equation}
\frac{1}{N}\sum_r F(r,t)=\left[F(k,t)\right]_{k=0},
\label{eq:I1}
\end{equation}
\begin{equation}
\frac{1}{N}\sum_r r^2 F(r,t) = -\left[\frac{d^2}{dk^2}F(k,t)\right]_{k=0}.
\label{eq:I2}
\end{equation}
Together, these give a result for a mean square displacement
\begin{equation}
\langle r^2 \rangle_F = \frac{\sum_r r^2 F(r,t)}{\sum_r F(r,t)},
\label{eq:I3}
\end{equation}
where $F$ is an arbitrary distribution.

 When the time variable is transformed to Laplace space rather than Fourier space,
$\Phi_Q(r,\omega)$ becomes $\int_0^\infty dt \exp(-\omega t) \Phi_Q(r,t)$.  The symbol $\omega$ previously used for the
Fourier variable is now used for the Laplace variable.  The solution of the Boltzmann equation is then exactly the same 
as previously, Eqns. \ref{eq:TAB} and \ref{eq:1En}, except that the Fourier variable $-i\omega$ becomes the Laplace variable $\omega$.
Using the fact that sums over $Q$ contains pairs $Q$ and $-Q$ with $v_{-Q}=-v_Q$,  which causes imaginary parts to vanish,
and using the notations $C_Q=\hbar\omega_Q dn_Q/dT$ and $\Lambda_Q=v_Q\tau_Q$, 
the functions that determine $\Delta T(k,\omega)$ and $\Delta E(k,\omega)$ are
\begin{equation}
A(k,\omega)=\frac{1}{N}\sum_Q \frac{\Delta E_Q^{\rm ext} (1+\omega\tau_Q)}{(1 +\omega\tau_Q)^2 +(k\Lambda_Q)^2},
\label{eq:2A}
\end{equation}
\begin{equation}
B(k,\omega) =\frac{1}{N}\sum_Q \frac{C_Q}{\tau_Q} \left[ 1-\frac{(1+\omega\tau_Q)}{(1 +\omega\tau_Q)^2 +(k\Lambda_Q)^2}\right],
\label{eq:2B}
\end{equation}
\begin{equation}
C(k,\omega)=\frac{1}{N}\sum_Q \frac{C_Q (1+\omega\tau_Q)}{(1 +\omega\tau_Q)^2 +(k\Lambda_Q)^2},
\label{eq:2C}
\end{equation}
\begin{equation}
D(k,\omega)=\frac{1}{N}\sum_Q \frac{\Delta E_Q^{\rm ext} \tau_Q (1+\omega\tau_Q)}{(1 +\omega\tau_Q)^2 +(k\Lambda_Q)^2}.
\label{eq:2D}
\end{equation}
Then recalling that $\Delta T(k,\omega)=A(k,\omega)/B(k,\omega)$, Eqs. \ref{eq:2A} and \ref{eq:2B} give Eq. 4 of Vermeersch
{\it et al.} in the V1 case.  They use $C_Q\rightarrow k_B$, the classical limit, and set $E_Q^{\rm ext}=E_0\rightarrow 1$ (which they
call a unit pulse).  The general answer for the integrated temperature rise of a pulse is
\begin{equation}
d_T(\omega)\equiv\frac{1}{N}\sum_r\Delta T(r,\omega) = \frac{1}{\omega} \frac{\sum_Q \frac{\Delta E_Q^{\rm ext}}{1+\omega\tau_Q}}
 {\sum_Q \frac{C_Q}{1+\omega\tau_Q}}.
\label{eq:DTsum}
\end{equation}
For the V1 pulse in the classical limit, this gives the total pulse temperature rise as $\sum_r k_B \Delta T(r,\omega)=E_0/\omega$.  Then 
the inverse Laplace transform gives the result of Ref. \onlinecite{VermeerschI2015}, $\sum_r k_B \Delta T(r,t)=E_0$.  However,
for other pulse forms of $E_Q^{\rm ext}$, there is no analytic inverse Laplace transform of Eq. \ref{eq:DTsum}.  Also, Vermeersch {\it et al.}
incorrectly identify $\sum_r k_B \Delta T(r,t)$ with the pulse energy $\sum_r \Delta E(r,t)$.  This identification is version {\bf (2)}
of Eq. \ref{eq:2versions}, but is inconsistent with version {\bf (1)} which they (sensibly) adopt as the preferred RTA approximation.
Fortunately, the erroneous identification of $\Delta E(r,t)$ with $k_B \Delta T(r,t)$ goes away in the V1 case when
summed over all $r$.  Using the correct RTA version
\begin{equation}
\Delta E(k,\omega)=k_B \frac{A(k,\omega)}{B(k,\omega)}C(k,\omega)+D(k,\omega),
\label{eq:}
\end{equation}
gives the result
\begin{equation}
\Delta E(k=0,\omega)=\frac{1}{\omega N}\sum_Q \Delta E_Q^{\rm ext}=\frac{E_0}{\omega}
\label{eq:}
\end{equation}
for all $\Delta E_Q^{\rm ext}$, not just the V1 case.
Doing the inverse Laplace transform then shows that the RTA in version {\bf (1)} correctly conserves the pulse energy in time,
\begin{equation}
\sum_r \Delta E(r,t)_{RTA,(1)}=E_0,
\label{eq:sumE}
\end{equation}
not just for the V1 pulse, but for arbitrary choice of $\Delta E_Q^{\rm ext}$.  This is not surprising; version {\bf (1)}
of Eq. \ref{eq:2versions} enforces overall energy conservation in RTA.  Version {\bf (2)} does not and does not obey
Eq. \ref{eq:sumE}.  The Vermeersch {\it et al.} identification
of $\Delta E(k,\omega)$ with $k_B \Delta T(k,\omega)$ is correct in the $k=0$ limit, but only for the V1 pulse, not for others.

	Now examine the mean square displacements, using Eq. \ref{eq:I2}.  For the case  $\Delta T(k,\omega)=A(k,\omega)/B(k,\omega)$,
the answer in Laplace space is
\begin{eqnarray}
n_T(\omega)&\equiv&\frac{1}{N}\sum_r r^2 \Delta T(r,\omega) = 2\frac{\sum_Q \frac{ \Delta E_Q^{\rm ext}\Lambda_Q^2}{(1+\omega\tau_Q)^3}}
{\sum_Q \frac{\omega C_Q}{(1+\omega\tau_Q)}} \nonumber \\
&+& 2\frac{\sum_Q \frac{\Delta E_Q^{\rm ext}}{(1+\omega\tau_Q)}} 
{ \left[ \sum_Q  \frac{\omega C_Q}{(1+\omega\tau_Q)}  \right]^2 } 
 \sum_Q   \frac{C_Q\Lambda_Q^2/\tau_Q}{(1+\omega\tau_Q)^3} 
\label{eq:DTr2sum}
\end{eqnarray}
For the V1 pulse this simplifies to
\begin{equation}
\frac{1}{N}\sum_r r^2 \Delta T(r,\omega) = \frac{2}{\omega^2}\frac{\sum_Q \frac{ C_Q \Lambda_Q^2/ \tau_Q}{(1+\omega\tau_Q)^2}}
{\sum_Q \frac{ C_Q}{(1+\omega\tau_Q)}},
\label{eq:}
\end{equation}
which is Eq. 6 of Vermeersch {\it et al.}.

The point of these calculations is that in principle we could do inverse Laplace transforms to get $n_T(t)$ from $n_T(\omega)$ 
and $d_T(t)$ from $d_T(\omega)$, and then take their ratio $n_T(t)/d_T(t)$ to get the 
mean square displacement $\langle r^2 (t)\rangle_{T}$ of the pulse temperature profile $\Delta T(r,t)$.  
Analytic inversions are not available.  Vermeersch et al. instead look at the limit of high $\omega \tau_Q$ which should give 
$\langle r^2 (t)\rangle_{T}$ at small times $t$, and also at the limit of low $\omega\tau_Q$ which should give
$\langle r^2 (t) \rangle_{T}$ at large times $t$.  

At large $\omega\tau_Q$, the $\omega$-dependences of $n_T(\omega)$ 
and $d_T(\omega)$ are 
\begin{eqnarray}
\frac{1}{N}\sum_r r^2 \Delta T(r,\omega) \rightarrow&& \frac{2}{\omega^3} 
\sum_Q [\Delta E_Q^{\rm ext}/\tau_Q]v_Q^2 / \sum_Q [ C_Q/\tau_Q] \nonumber \\
&&+ {\mathcal O}\left(\frac{1}{\omega^4}\right), 
\label{eq:}
\end{eqnarray}
\begin{eqnarray}
\frac{1}{N}\sum_r \Delta T(r,\omega) \rightarrow && \frac{1}{\omega}\sum_Q [\Delta E_Q^{\rm ext}/\tau_Q] / \sum_Q [ C_Q/\tau_Q]   
\nonumber \\
&&+ {\mathcal O}\left( \frac{1}{\omega^2} \right). 
\label{eq:}
\end{eqnarray}
The inverse Laplace transform of $2/\omega^3$ is $t^2$, and of $1/\omega$ is $1$.  Then we get 
ballistic behavior in the short time limit,
\begin{equation}
\langle r^2 \rangle_T \rightarrow \langle v^2 \rangle_T t^2
\label{eq:}
\end{equation}
where the mean square velocity of the temperature pulse is
\begin{equation}
\langle v^2 \rangle_T \rightarrow \sum_Q [\Delta E_Q^{\rm ext}/\tau_Q] v_Q^2 / \sum_Q [\Delta E_Q^{\rm ext}/\tau_Q].
\label{eq:v2T}
\end{equation}
This agrees with Eq. 8 of Vermeersch {\it et al.} in the V1 case.  It also resembles our result for the collisionless 
energy pulse propagation,
namely Eqs. \ref{eq:Ext} or \ref{eq:ExtB} plus Eq. \ref{eq:xbarcont}.  However, in contrast with the collisionless limit,
there is an extra $Q$ and $T$-dependent factor
$1/\tau_Q$ in the weights of $\langle v^2 \rangle$ in Eq. \ref{eq:v2T}.  This is clearly wrong; collisions cannot alter
the free phonon ballistic propagation velocity at short times when few or no collisions occur.

At small $\omega\tau_Q$, the $\omega$-dependences of $n_T(\omega)$ 
and $d_T(\omega)$ are 
\begin{eqnarray}
\frac{1}{N}\sum_r r^2 \Delta T(r,\omega) \rightarrow&& \frac{2}{\omega^2} 
\frac{[\sum_Q \Delta E_Q^{\rm ext}][ \sum_Q C_Q \Lambda_Q^2/\tau_Q]}  {[\sum_Q  C_Q]^2} \nonumber \\
&&+ {\mathcal O}\left(\frac{1}{\omega}\right), 
\label{eq:}
\end{eqnarray}
\begin{eqnarray}
\frac{1}{N}\sum_r \Delta T(r,\omega) \rightarrow && \frac{1}{\omega}[\sum_Q \Delta E_Q^{\rm ext}] / [\sum_Q  C_Q]   
\nonumber \\
&&+ {\mathcal O} (1). 
\label{eq:}
\end{eqnarray}
The inverse Laplace transform of $2/\omega^2$ is $2t$, and of $1/\omega$ is $1$.  Then we get 
diffusive behavior in the long time limit,
\begin{eqnarray}
&&\langle r^2 \rangle_T \rightarrow 2 D t, \ \ {\rm where} \ \ D=\kappa/C \nonumber \\
&& \kappa =  \sum_Q C_Q \Lambda_Q^2/\tau_Q \ \ {\rm and} \ \ C=\sum_Q C_Q.
\label{eq:}
\end{eqnarray}
This result is independent of how the pulse inserts energy $\Delta E_Q^{\rm ext}$, and the diffusivity $D$
has the correct macroscopic value.  This contradicts the argument that pure diffusion isn't in the nonlocal PBE.

Why would the nonlocal PBE give correct long time diffusion but incorrect short time ballistic?  The answer, we think,
is that version {\bf (1)} of RTA, namely energy conservation $\sum_Q \hbar\omega_Q\Phi_Q/\tau_Q$ doesn't work
when $\tau_Q\rightarrow\infty$.

\bibliography{Bib}

\end{document}


\title{ \ \ \ \ \ \ \ \ \ \ Supplemental Material for
\newline
\newline
Nonlocal Phonon Heat Transport Seen in 1-d Pulses}

\author{Philip B. Allen}
\email{philip.allen@stonybrook.edu}
\affiliation{Department of Physics and Astronomy, Stony Brook University, Stony Brook, NY 11794-3800, USA}

\author{Nhat A. Nghiem}
\affiliation{Department of Physics and Astronomy, Stony Brook University, Stony Brook, NY 11794-3800, USA}

\date{\today}




\maketitle

In the first section, some mathematics of disordered harmonic systems is used to show that
thermal disorder in a harmonic chain will not erase the details of atomistic pulse energy, as shown
in Fig. 3 of the full text.  In the second section, formulas are given for the five functions used to compute
the pulse energy in $(k,\omega)$-space, and the $Q$-integration is described.  In the last section,
some details of the numerical Fourier transform to $(x,t)$-space are given.  When equations from
the full text are cited, they are indicated as, for example, Eq. T6.

\section{Classical harmonic disordered lattice}

The most general harmonic linear chain is described by
\begin{equation}
\mathcal{H}=\frac{1}{2} \sum_\ell \dot{w}_\ell^2 +\frac{1}{2} \sum_\ell \Omega_{\ell\ell^\prime} w_\ell w_{\ell^\prime},
\label{eq:A1}
\end{equation}
where $w_\ell = \sqrt{M_\ell} u_\ell$ and $\Omega_{\ell\ell^\prime}=K_{\ell,\ell^\prime}/\sqrt{M_\ell M_{\ell^\prime}}$.
The spring constant $K_{\ell\ell^\prime}$ is $\partial^2 U_{\rm PE}/\partial u_\ell \partial u_{\ell^\prime}$.
Vibrational eigenstates $|\mu\rangle$ are solutions of $\hat{\Omega}|\mu\rangle = \omega_\mu^2 |\mu\rangle$.
The general solution of Newton's laws is described by $N$ amplitudes $A_\mu$ and phases $\phi_\mu$,
\begin{equation}
w_\ell(t)=\frac{1}{\sqrt{N}}\sum_\mu A_\mu \mu_\ell \cos(\omega_\mu t - \phi_\mu), \ {\rm where} \ \mu_\ell = 
\langle \ell|\mu\rangle.
\label{eq:A2}
\end{equation}
The normal mode coordinates $\mu_\ell=\langle \ell | \mu \rangle$ are defined to be real numbers.  For ordered chains of atoms, the
simpler convention is to use complex numbers $\langle \ell | Q\rangle=\exp(iQ\ell a)/\sqrt{N}$, but real numbers (cosines and sines)
also work.

The normal modes are orthonormal,
\begin{equation}
\langle \mu | \mu^\prime \rangle = \sum_\ell \langle \mu |\ell\ \rangle \langle \ell | \mu^\prime \rangle 
=\frac{1}{N} \sum_\ell \mu_\ell \mu_\ell^\prime = \delta_{\mu\mu^\prime}. 
\label{eq:A3}
\end{equation}
From this it is easy to show that
\begin{eqnarray}
{\rm KE} &=& \frac{1}{2}\sum_\mu \omega_\mu^2 A_\mu^2 \sin^2(\omega_\mu t - \phi_\mu), \nonumber \\
{\rm PE} &=& \frac{1}{2}\sum_\mu \omega_\mu^2 A_\mu^2 \cos^2(\omega_\mu t - \phi_\mu),
\label{eq:A4}
\end{eqnarray}
and therefore the total energy is 
\begin{equation}
E_{\rm tot}=\sum_\mu E_\mu, \ \ {\rm where} \ \ E_\mu = \frac{1}{2} \omega_\mu^2 A_\mu^2.
\label{eq:A5}
\end{equation}
This is the generalization to lattices (whether ordered or disordered), of Eq. T8 for ordered lattices.
It is also easy to derive the formula
\begin{equation}
A_\mu e^{i(\phi_\mu - \omega_\mu t)} = \frac{1}{\sqrt{N}} \sum_\ell \sqrt{M_\ell} \left( u_\ell(t) + i\frac{v_\ell(t)}{\omega_\mu} 
\right) \mu_\ell.
\label{eq:A6}
\end{equation}
This is the generalization of Eq. T6.
From this formula, one can extract the amplitude $A_\mu$ and phase $\phi_\mu$ of any normal mode $\mu$ if the
positions $u_\ell$ and velocities $v_\ell$ are known at any particular time $t$.  For example, consider a V1 pulse
inserted at zero temperature.  Using Eq. \ref{eq:A6} at $t=0$, the modal energy is
\begin{eqnarray}
E_\mu^{{\rm V1},T=0} = \frac{1}{2} \omega_\mu^2 A_\mu^2 &=& \frac{M_{\ell=0}}{2N} \Delta v_{\ell=0}^2 \mu_{\ell=0}^2 
\nonumber \\
&=& E_{\rm pulse}^{\rm V1} \frac{\langle {\ell=0} |\mu\rangle\langle\mu | {\ell=0} \rangle}{N} \nonumber \\
E_{\rm pulse}^{\rm V1}&=& \sum_\mu E_\mu^{{\rm V1},T=0} = \frac{M_0 \Delta v_0^2}{2}.
\label{eq:A7}
\end{eqnarray}

From these results, we can make a tedious proof (for the V1 case) 
that when temperature is not zero, the ensemble average energy
in each mode $\mu$ continues to obey Eq. \ref{eq:A7}.  This means that the ensemble average pulse shape is
unaffected by temperature, for both ordered and disordered chains, agreeing with our simulation results in Fig. 3
for the ordered chain.  For any particular member of the ensemble, the mode $\mu$ has energy 
$\frac{1}{2} \omega_\mu^2 A_\mu^2$ before pulse insertion, and $\frac{1}{2} \omega_\mu^2 (A_\mu+\Delta A_\mu)^2$
after.  Using Eq. \ref{eq:A6}, this can be written as
\begin{eqnarray}
E_\mu^{\rm before} &=& [f_\mu + i g_\mu v_0][f_\mu^\ast -ig_\mu v_0] \nonumber \\
E_\mu^{\rm after} &=& [f_\mu + i g_\mu (v_0+\Delta v_0)][f_\mu^\ast -ig_\mu (v_0+\Delta v_0)], \nonumber \\
\label{eq:A8}
\end{eqnarray}
where 
\begin{equation}
f_\mu = \sqrt{\frac{1}{2N}} \left[\sum_\ell^{\ne 0}\sqrt{M_\ell}(\omega_\mu u_\ell + i v_\ell)\mu_\ell 
+ \sqrt{M_0} \omega_\mu u_0 \mu_0 \right]
\label{eq:A9}
\end{equation}
\begin{equation}
{\rm and} \ \ g_\mu = \sqrt{\frac{M_0}{2N}} \mu_0.
\label{eq:A10}
\end{equation}
The point of Eq. \ref{eq:A8} is to separate the coordinate $v_0$ or $v_0+\Delta v_0$ from the remaining coordinates
$u_\ell, v_\ell$ which are collected in $f_\mu$.  This enables ensemble averages to be done using the fact that
$\langle f_\mu v_0\rangle_T = \langle f_\mu \rangle_T \langle v_0\rangle_T = 0$ and 
$\langle f_\mu \Delta v_0\rangle_T = \langle f_\mu \rangle_T \langle \Delta v_0\rangle_T = 0$.
The null values follow from the fact that $\langle f_\mu \rangle_T =0$.
Note that $\langle \Delta v_0\rangle_T \ne 0$; $\Delta v_0$ is chosen to make the V1 pulse have a fixed energy 
$E_{\rm pulse}$, for every member of the ensemble.  This requires
\begin{equation}
\frac{1}{2}M_0 [(v_0 +\Delta v_0)^2 - v_0^2 ] = E_{\rm pulse}^{\rm V1}
\label{eq:A11}
\end{equation}
whether ensemble-averaged or not.  This is the general version of the $T=0$ form in Eq. \ref{eq:A7} which has $v_0=0$.
From Eq. \ref{eq:A8} we get
\begin{equation}
\Delta E_\mu^{\rm V1} = g_\mu^2 [(v_0 + \Delta v _0)^2-v_0^2] -i(f_\mu - f_\mu^\ast)g_\mu \Delta v_0.
\label{eq:A12}
\end{equation}
The ensemble average is 
\begin{equation}
\langle \Delta E_\mu^{\rm V1} \rangle_T = g_\mu^2 \langle [(v_0 + \Delta v _0)^2-v_0^2]\rangle_T.
\label{eq:A13}
\end{equation}
Finally, using Eqs. \ref{eq:A7}, \ref{eq:A10},  and \ref{eq:A11}, we get
\begin{equation}
\langle \Delta E_\mu^{\rm V1} \rangle_T = E_{\rm pulse}^{\rm V1} \frac{\mu_0^2}{N}, \  {\rm and} \ 
\sum_\mu \langle \Delta E_\mu^{\rm V1} \rangle_T = E_{\rm pulse}^{\rm V1}.
\label{eq:}
\end{equation}
This proof is sufficiently tedious that the corresponding proofs for V2 and D pulses have not been checked.
The results of Fig. 3 indicate that probably all pulses on harmonic chains have shapes independent of $T$
after ensemble averaging.



\section{Numerical computation of $\Delta E(k,\omega)$}

The local energy in Fourier space is
\begin{equation}
\Delta E(k,\omega)=\sum_Q \hbar\omega_Q \Phi_Q(k,\omega), 
\label{eq:}
\end{equation}
where $\Phi(k,\omega)$ was derived from Boltzmann theory in relaxation time approximation (RTA):
\begin{equation}
\Phi_Q(k,\omega) =
\frac{-i(kv_Q-\omega)\frac{\partial n_Q}{\partial T}\Delta T(k,\omega) + \Delta E_Q^{\rm ext}/\hbar\omega_Q}
{1/\tau_Q+i(kv_Q-\omega)}.
\label{eq:RTAsol}
\end{equation}
The extra equation needed to eliminate the unknown $\Delta T(k,\omega)$ has two forms,
\begin{equation}
\bf{(1)} \ \sum_Q \frac{\hbar\omega_Q \Phi_{Q,{\rm RTA}}}{\tau_Q}=0; \ {\rm or} \ {\bf (2)} \ 
\sum_Q \hbar\omega_Q \Phi_{Q,{\rm RTA}} =0. 
\label{eq:2versions}
\end{equation}

In the appendix of the main paper, these results are used to derive the following
\begin{equation}
\Delta E_{(1)}(k,\omega) = C(k,\omega)\frac{A(k,\omega)}{B(k,\omega)} + D(k,\omega),
\label{eq:1En}
\end{equation}
and
\begin{equation}
\Delta E_{(2)}(k,\omega)=\frac{D(k,\omega)}{F(k,\omega)}, 
\label{eq:}
\end{equation}

Equations for $A(k,\omega)$ , $\ldots$, $F(k\omega)$ are in that appendix.  Here
we convert them to dimensionless functions
in dimensionless variables $x=kv_M/\omega_M=ka/2$, $y=\omega/\omega_M$, and $z=Qa/2$.
The definitions are
\begin{eqnarray}
A(k,\omega)&=&\frac{8E_0}{3N}a(x,y),  \ \nonumber \\
B(k,\omega)&=& \omega_M b(x,y),  \ \nonumber \\
C(k,\omega)&=&c(x,y), \ \nonumber \\
D(k,\omega)&=&\frac{8E_0}{3N\omega_M}d(x,y), \ \nonumber \\
F(k,\omega)&=&f(x,y) 
\label{eq:abcdf}
\end{eqnarray}
The  local energy density in the two versions is
\begin{equation}
\Delta E_{(1)}(x,y) = \left(\frac{8E_0/3N}{\omega_M}\right) \left[\frac{a(x,y)}{b(x,y)}c(x,y)+d(x,y)\right].
\label{eq:dd}
\end{equation}
\begin{equation}
\Delta E_{(2)}(x,y)= \left(\frac{8E_0/3N}{\omega_M}\right) \frac{d(x,y)}{f(x,y)}.
\label{eq:2DT}
\end{equation}
The dimensionless functions of $(x,y)$ are
\begin{equation}
a(x,y)=\frac{1}{\pi} \int_{-\pi/2}^{\pi/2} dz \frac{\sin^4 z}{1-\frac{i}{\epsilon} \frac{\cos z}{\sin^2 z}\left(y-x \cos z \ {\rm sign}(z) \right)},
\label{eq:aa}
\end{equation}
\begin{equation}
b(x,y)=\frac{1}{\pi} \int_{-\pi/2}^{\pi/2} dz \frac{-i \left(y-x \cos z \ {\rm sign}(z) \right) } { 1-\frac{i}{\epsilon} 
\frac{\cos z}{\sin^2 z} \left(y-x \cos z \ {\rm sign}(z) \right) },
\label{eq:bb}
\end{equation}
\begin{equation}
c(x,y)=\frac{1}{\pi} \int_{-\pi/2}^{\pi/2} dz \frac{1}{1-\frac{i}{\epsilon} \frac{\cos z}{\sin^2 z}\left(y-x \cos z \ {\rm sign}(z) \right)},
\label{eq:cc}
\end{equation}
\begin{equation}
d(x,y)=\frac{1}{\epsilon\pi} \int_{-\pi/2}^{\pi/2} dz \frac{\cos z \sin^2 z}{1-\frac{i}{\epsilon}\frac{\cos z}{\sin^2 z}
\left(y-x \cos z \ {\rm sign}(z) \right)}.
\label{eq:dd}
\end{equation}
\begin{equation}
f(x,y)=\frac{1}{\pi} \int_{-\pi/2}^{\pi/2} dz \frac{-i \left(y-x \cos z \ {\rm sign}(z) \right) } { \frac{\epsilon\sin^2 z}{\cos z} 
-i \left(y-x \cos z \ {\rm sign}(z) \right) }.
\label{eq:ff}
\end{equation}

For given $k,\omega$, the $z=Qa/2$ integrations often converge fairly slowly.  
For 80 $Q$-points (from 0 to $\pi/a$) the error is usually about $2-4\times 10^{-3}$,
and for 640 $Q$-points the error is $\sim 2\times 10^{-6}$.  Final calculations of
these functions of $(x,y)$ used 1600 $Q$-points.

\section{Numerical computation of $\Delta E(r,t)$}

The Fourier transform needed is
\begin{equation}
\Delta E(r,t)=\frac{16E_0}{3N} \int_{-\infty}^{\infty} \frac{dk}{2\pi} g(k,t) e^{i(2r/a) x}, \ {\rm where}
\label{eq:FTk}
\end{equation}
\begin{equation}
g(k,t)=\int_{-\infty}^\infty \frac{dy}{2\pi} h(x,y) e^{-i(\omega_M t)y},
\label{eq:FTw}
\end{equation}
and $h=(a/b)c+d$ or $h=d/f$ in cases (1) and (2) respectively.
Since $h(x,-y)=h^\ast(x,y)$, Eq. \ref{eq:FTw} can be simplified to
\begin{equation}
g(x,t) = \frac{1}{\pi} \int_0^\infty dy \mathcal{R}{\rm e}\left\{ e^{-i(\omega_M t)y} h(x,y) \right\}.
\label{eq:gy}
\end{equation}
Since $g(x,t)$ is even in $x$, Eq. \ref{eq:FTk} can be simplified to
\begin{equation} 
\Delta E(r,t)=\frac{16E_0}{3\pi} \int_0^\infty dx \cos[(2r/a)x] g(x,t)
\label{eq:kTe}
\end{equation}
This shows that $\Delta E(r,t)$ is real, as physically required.

We need to evaluate $\Delta E(r,t)$ for values of $r$ equal to $na$, $n=0,1, ... ,40$ and
for $t=80/\omega_M=40$ in units $\sqrt(M/K)$.  The factor $h(x,y)$ varies over a wide range of $x,y$ or $k,\omega$,
but does not vary too rapidly within the relevant decades of $(x,y)$.   However, the
Fourier factors $\exp(ikr-i\omega t)$ can vary enormously over a decade of $x$ or $y$.
A way to integrate efficiently is to use logarithmic grids for $x$ and $y$, with 
$n_g$ points per decade, and $n_d \sim$ 10 decades.  We used $n_g=500$ for the $y=\omega/\omega_M$ integral  
and integrated $y$ from $10^{-8}$ to $10^5$.  For the $x=ka/2$ integrals we used $n_g=40$ points per
decade, and integrated from $x=10^{-7}$ to $x=10^4$.
Between grid points, $h(x,y)$ is interpolated linearly,
and the integration is done exactly with linearized $h$ .  For example,
consider the $y$ integration of Eq. \ref{eq:gy},
\begin{equation}
g(x,t) = \sum_{n=0}^{n_d \times n_g} \Delta g_n
\label{eq:}
\end{equation}
\begin{equation}
\Delta g_n\equiv \frac{1}{\pi} \mathcal{R}{\rm e} \int_{y_n}^{y_{n+1}} dy e^{-i(\omega_M t)y} h(x,y),
\label{eq:glin}
\end{equation}
with $y_{n+1}=r_y y_n$ and $(r_y)^{500}=10$.
In the interval $(y_n,y_{n+1})$, approximate $h(x,y)$ by a linear interpolation:
\begin{equation}
 h(x,y_n)+(y-y_n)\left[\frac{h(x,y_{n+1})-h(x,y_n)}{y_{n+1}-y_n} \right].
\label{eq:glin}
\end{equation}
The linearized integral in Eq. \ref{eq:glin} is
\begin{eqnarray}
&& \ \ \ \ \ \pi \Delta g_n(x=ka/2,t) \nonumber \\
&&\approx \frac{1}{-i\omega_M t}
\left[e^{-i(\omega_M t)y_{n+1}}h(x,y_{n+1}) - e^{-i(\omega_M t)y_{n}}h(x,y_{n}) \right] \nonumber \\
&&-\left( \frac{1}{-i\omega_M t} \right)^2 \left[e^{-i(\omega_M t)y_{n+1}} - e^{-i(\omega_M t)y_{n}} \right] \nonumber \\
&&\times\frac{h(x,y_{n+1})-h(x,y_n)}{y_{n+1}-y_n}.
\label{eq:gn}
\end{eqnarray}
%



\bibliography{Bib}